\documentstyle[12pt,epsf]{article}

\begin{document} 

\begin{titlepage}

\hrule 
\leftline{}
\leftline{Chiba Univ. Preprint
          \hfill   \hbox{\bf CHIBA-EP-128}}
\leftline{\hfill   \hbox{hep-th/0105268}}
\leftline{\hfill   \hbox{May 2001}}
\vskip 5pt
\hrule 
\vskip 1.0cm
\centerline{\large\bf
  The Most General and Renormalizable Maximal Abelian Gauge
}

\vskip 1cm

\centerline{{\bf 
Toru Shinohara,${}^{1,\dagger}$
Takahito Imai${}^{1}$ and
Kei-Ichi Kondo${}^{1,2,\ddagger}$
}}
\vskip 1cm
\begin{description}
\item[]{\it 
$^1$ Graduate School of Science and Technology,
  Chiba University, Chiba 263-8522, Japan
  }
\item[]{\it \centerline{ 
$^2$ Department of Physics, Faculty of Science, 
Chiba University,  Chiba 263-8522, Japan}
  }
\end{description}

\centerline{{\bf Abstract}}

We construct the most general gauge fixing and the associated Faddeev-Popov 
ghost term
for the $SU(2)$ Yang-Mills theory,
which leaves the global $U(1)$ gauge symmetry intact
(i.e., the most general Maximal Abelian gauge).  We show that the most 
general form involves eleven independent gauge parameters.
Then we require various symmetries which help to reduce the number of 
independent parameters for obtaining the simpler form.
In the simplest case, the off-diagonal part of the gauge fixing term 
obtained in this way
is identical to the modified maximal Abelian gauge term with two gauge 
parameters which was
proposed in the previous paper from the viewpoint of renormalizability.
  In this case, moreover, we calculate the beta function,
anomalous dimensions of all fields
and renormalization group functions of all gauge parameters
in perturbation theory to one-loop order.
We also discuss the implication of these results to obtain information 
on low-energy physics of QCD.


\vskip 0.2cm
\hrule  
${}^\dagger$ 
  E-mail: {\tt sinohara@cuphd.nd.chiba-u.ac.jp},
          {\tt sinohara@graduate.chiba-u.jp}
\par 
${}^\ddagger$ 
  E-mail: {\tt kondo@cuphd.nd.chiba-u.ac.jp},
          {\tt kondok@faculty.chiba-u.jp}


\vskip 0.5cm  


\end{titlepage}



\section{Introduction} %
\par
The gauge fixing is an indispensable procedure
in quantizing the continuum gauge theory.
It is believed that the physically meaningful results do not
depend on the gauge fixing condition.
Therefore we can adopt any favorite gauge fixing condition
for obtaining physical quantities.
We often adopt the Lorentz gauge in order to simplify the calculation.
In this gauge, the Lorentz covariance is explicitly preserved.
Especially, the Landau gauge is very efficient to simplify the calculations 
in the gauge theory.

In this paper, nevertheless, we consider the maximal Abelian (MA) gauge.
This gauge is very useful to clarify the low energy physics
  of quantum chromodynamics (QCD), since
in the low energy region of QCD, the Abelian projection
procedure \cite{tHooft81} or a hypothesis of Abelian dominance \cite{EI82}
has been justified by the recent research mainly based on numerical
simulations \cite{SY90}.
  Especially, quark confinement can be explained
based on the dual superconductor picture of QCD vacuum \cite{dsc} at least 
qualitatively.
Dual superconductivity of QCD is expected to be described
by the dual Ginzburg-Landau (DGL) theory.
However, the DGL theory is an Abelian gauge theory, while QCD is
an $SU(3)$ non-Abelian gauge theory.
Therefore, for the dual superconductor picture to be responsible
for the low energy physics in QCD such as quark confinement and spontaneous 
breakdown of chiral symmetry, the Abelian
projection procedure turns out to be useful.
Thus we expect that the maximal Abelian gauge \cite{QR98} is the most 
useful gauge
for describing the low energy physics of QCD.

In a series of papers\cite{KondoI,KondoII,KondoIII,KS00a,KS00b,Shinohara01a},
we have tried to give an analytical framework
which enables us to explain the  Abelian dominance in QCD under the MA gauge
from the viewpoint of renormalizability.
The MA gauge  is a nonlinear gauge fixing condition, in sharp contrast with 
the conventional gauge fixing of the Lorentz type which is a linear gauge.
Due to this non-linearity, we must introduce the quartic ghost--antighost
self-interaction to maintain the renormalizability.
The {\it modified} MA gauge fixing term \cite{KondoII,KS00a} was devised to 
incorporate such a self-interaction term in a natural way.
We have pointed out a possibility of dynamical
mass generation of off-diagonal gluons and off-diagonal ghosts
due to the ghost--antighost condensation.
The fact that the off-diagonal fields become massive while the diagonal 
fields remain massless or have smaller masses gives an analytical 
explanation of Abelian dominance
in the low energy region.

In spite of such an advantage in the low energy region of QCD, the MA gauge is
rarely adopted in contrast with the Lorentz gauge.
One of the reasons is that the calculation in the MA gauge is very
complicated because of the nonlinearity of MA gauge fixing condition.
Another reason is that the MA gauge partially fixes the non-Abelian
gauge symmetry leaving the residual $U(1)^{N-1}$ gauge symmetry
so that the global color symmetry is partially broken
and we must distinguish diagonal components and off-diagonal
components of fields in the MA gauge.
Therefore, the detailed investigation of the MA gauge has not yet been 
performed even in the perturbative level except for pioneering works 
\cite{MLP85}.
In this paper, therefore, we present perturbative results, i.e., 
calculations of the beta-function, anomalous dimensions of fields,
and renormalization group (RG) function of gauge parameters in the MA gauge.

Before performing concrete calculations, we obtain the most general form
of $SU(2)$ Yang-Mills action in the MA gauge.
We show that the most general form of the MA gauge involves eleven gauge 
parameters.
  We classify the gauge parameter space
from the viewpoint of symmetries.
A detailed consideration of such a gauge fixing term in the case
of $SU(2)$ has already been attempted by Min, Lee and Pac\cite{MLP85}
or Hata and Niigata\cite{HN93} and some of RG functions
of gauge parameters were  calculated there.
However, in this paper, we give thorough analyses  of
the symmetries imposed on the possible action
in the MA gauge.
In the simplest case, the off-diagonal part of the gauge fixing terms 
obtained in this way
is identical to the modified maximal Abelian gauge terms with two gauge 
parameters
proposed in the previous paper from the viewpoint of renormalizability.
Moreover, we calculate anomalous dimensions of all fields
and RG functions of gauge parameters appearing in our action
to one-loop order of perturbation theory in the scheme of the dimensional 
regularization.

Even though our main interest lies  in the investigation of the low
energy physics of QCD and the perturbative approach is not valid there, the 
perturbative calculations are the first step toward the non-perturbative 
studies of the low energy physics governed by the strong coupling dynamics.
This is because  the high energy behavior
  could be related to the low energy one by renormalization group equation 
and analyticity.\cite{OZ80}
In fact, the anomalous dimensions and RG functions
calculated by perturbative method give indispensable ingredients for the 
nonperturbative approaches, for instance, the truncated coupled 
Dyson-Schwinger equations, superconvergence relations\cite{CEP139,AS01}
and numerical simulations on a lattice\cite{BCGMP}.

This paper is organized as follows.
In section 2, we give a general consideration on the renormalizable gauge 
fixing and FP ghost term respecting the global $U(1)$ gauge symmetry in 
the $SU(2)$ non-Abelian gauge theory.
By  taking account of the symmetries, we can fix some of the
parameters without spoiling the renormalizability.
Then we restrict our consideration to a fixed parameter subspace.
It is possible to choose a minimal version of the maximal Abelian gauge
by restricting the parameter space to three independent parameters.
In section 3, we calculate quantum corrections to
all the remaining parameter in the minimal 
choice of the most general MA gauge,
although some of the anomalous dimensions
have already been obtained in the previous papers\cite{KS00b,Shinohara01a}.
The renormalizability of the  modified  MA gauge is confirmed to one-loop 
order of perturbation theory.
We give the conclusion and discussion in the final section.
In Appendix~\ref{sec:rescale}, we discuss the rescaling of the fields
preserving BRST transformation and its connection to the renormalization.

\section{The most general gauge fixing terms} %
\label{sec:The most general gauge fixing terms}
In this section, we construct the most general gauge fixing (GF)
and the associated Faddeev-Popov (FP) ghost term
for the Maximal Abelian gauge in the $SU(2)$ Yang-Mills theory.
Note that we require only the global $U(1)$ symmetry
for the gauge fixing term, not the {\it global} $SU(2)$ gauge symmetry.
The most general FP+GF term is obtained in the BRST exact form,
\begin{equation}
S_{\rm GF+FP}
 =-i\int d^4x\mbox{\boldmath$\delta$}_{\rm B} G ,
\label{eq:GF+FP}
\end{equation}
where $G$
is a  functional of
gluons ${\cal A}_{\mu}^A=(A_\mu^a,a_\mu)$,
ghosts ${\cal C}^A=(C^a,C^3)$,
antighosts $\bar{\cal C}^A=(\bar C^a,\bar C^3)$ 
and Nakanishi-Lautrup fields ${\cal B}^A=(B^a,B^3)$ with $a=1,2$.
\par
First, the functional $G$ must satisfy the following requirements.
\begin{enumerate}
\item[1)]
 The functional $G$ must be of mass dimension 3.
Since ${\cal A}_{\mu}^A=(A_\mu^a,a_\mu)$,
${\cal C}^A=(C^a,C^3)$,
$\bar{\cal C}^A=(\bar C^a,\bar C^3)$ and
${\cal B}^A=(B^a,B^3)$ has respectively the mass dimension  1, 1, 1 and 2,  a monomial in the functional $G$ consists of at most three fields.

\item[2)]
The functional $G$ must have the global $U(1)$ symmetry.  

\item[3)]
The functional $G$ must have the ghost number $-1$.
Note that ${\cal A}_{\mu}^A=(A_\mu^a,a_\mu)$,
${\cal C}^A=(C^a,C^3)$,
$\bar{\cal C}^A=(\bar C^a,\bar C^3)$ and
${\cal B}^A=(B^a,B^3)$ has the ghost number 0, 1, $-1$ and 0, respectively.

\end{enumerate}
From the above requirements 1) and 2), the possible form of the monomials in $G$ can be classified into seven groups:
\begin{equation}
  \delta^{ab}X^a Y^b Z^3,
\quad
  \epsilon^{ab3}X^a Y^b Z^3, 
\quad
  X^3 Y^3 Z^3 ,
\end{equation}
\begin{equation}
  \delta^{ab} X^a Y^a,
\quad 
  \epsilon^{ab} X^a Y^b,
\quad
  X^3 Y^3 ,
\end{equation}
\begin{equation}
  X^3 .
\end{equation}
Taking account of the fact that the functional $G$ is of the form, 
\begin{equation}
 G=\bar{C} \Phi ,
\end{equation}
apart from the index, we find that one of $X$, $Y$ and $Z$ must be an antighost $\bar{\cal C}^A=(\bar C^a,\bar C^3)$ and that $\Phi$ must be of dimension 2 and of ghost number zero from the requirement 3).  

\par
Second, we consider the global $SU(2)$ symmetry which is broken by the MA gauge fixing. 
The invariants under the global $SU(2)$ rotation are 
\begin{equation}
  \epsilon^{ABC} X^A Y^B Z^C, \quad \delta^{AB} X^A X^B .
\end{equation}
Therefore, the three groups of the seven groups belong to this type:
\begin{equation}
  \epsilon^{ab3}X^a Y^b Z^3, 
\quad
   \delta^{ab} X^a Y^b,
\quad
  X^3 Y^3 .
  \label{eq:1stgroup}
\end{equation}
The remaining four groups,
\begin{equation}
  \delta^{ab}X^a Y^b Z^3,
\quad
  X^3 Y^3 Z^3 ,
\quad
  \epsilon^{ab} X^a Y^b,
\quad
  X^3 ,
  \label{eq:2ndgroup}
\end{equation}
 are incompatible with the global $SU(2)$ symmetry if they exist in the functional $G$.
They are called the {\it exceptional} terms.  
Thus, the possible form of the functional is rewritten as
\begin{equation}
S_{\rm GF+FP}
 =-i\int d^4x\mbox{\boldmath$\delta$}_{\rm B}
   \left(G^{(a)}+G^{(i)}+G^{\rm ex}\right),
\label{eq:S_GF+FP}
\end{equation}
where
we have decomposed the terms belonging to the first group (\ref{eq:1stgroup}) into two
 functionals~$G^{(a)}$ and $G^{(i)}$ according to their forms,
\begin{equation}
  G^{(a)} = \bar{C}^a \Phi^a ,
\quad
  G^{(i)} = \bar{C}^3 \Psi^3 ,
\end{equation}
and
$G^{\rm ex}$ denotes the exceptional terms of the form,
\begin{equation} 
  G^{\rm ex} = \bar{C}^a \Phi'^a + \bar{C}^3 \Psi'^3 .
\end{equation}
\par
The first functional~$G^{(a)}$ plays the role of partially fixing
the $SU(2)$ gauge symmetry to $U(1)$.
The possible form of monomials in $G^{(a)}$ is either 
$\epsilon^{ab3}\bar{C}^aY^b Z^3$
or
$\delta^{ab}\bar{C}^aY^b$.
It is easy to see that the possible choices are given as
$
 \epsilon^{ab3}\bar{C}^aY^b Z^3
 \sim \epsilon^{ab3}\bar{C}^a (C^b \bar{C}^3, \bar{C}^b C^3, A_\mu^b a_\mu) 
$
and
$
 \delta^{ab} \bar{C}^aY^b 
\sim \delta^{ab} \bar{C}^a (B^b, \partial^\mu A_\mu^b)  .
$
Thus the most general form of $G^{(a)}$ is given by
\begin{equation}
G^{(a)}
 :=\bar C^a\left[
    (\partial^\mu\delta^{ab}-\xi g\epsilon^{ab}a^\mu)A_\mu^b
     +\frac\alpha2B^a
     -i\frac\zeta2g\epsilon^{ab}\bar C^bC^3
     +i\eta g\epsilon^{ab}\bar C^3C^b
    \right].
\label{eq:G^a}
\end{equation}
It turns out that the off-diagonal component of the Nakanishi-Lautrup
field $B^a$ is generated from this functional after performing 
the BRST transformation explicitly.
\par
By making use of the anti-BRST transformation, this functional is recast  into
\begin{eqnarray}
G^{(a)}
 &\equiv&
   -\bar{\mbox{\boldmath$\delta$}}_{\rm B}
    \left[\frac12A^{\mu a}A_\mu^a
          -\frac\zeta2iC^a\bar C^a\right]
   \nonumber\\
 & &+\bar C^a\left[
     i(1-\xi)g\epsilon^{ab}a^\mu A_\mu^a
     +\frac{\alpha-\zeta}2B^a
     +i\eta g\epsilon^{ab}\bar C^3C^b
    \right] .
\label{eq:G'}
\end{eqnarray}
The first term of the right hand side of (\ref{eq:G'})
is both BRST and anti-BRST exact, and give rise to
the  modified  MA gauge fixing term
proposed in the previous papers\cite{KondoII,KS00a}.
After performing the BRST transformation, we obtain
\begin{eqnarray}
S^{(a)}
 &:=&
    -i\int d^4x\mbox{\boldmath$\delta$}_{\rm B}G^{(a)}
    \nonumber\\
 &=&\int d^4x\biggl\{
    \frac{\alpha}{2}B^aB^a
    +
    B^aD^{\xi\mu} A_\mu^a
    -
    ig\zeta\epsilon^{ab}
    B^a\bar{C}^bC^3
    +
    ig\eta\epsilon^{ab}B^a\bar{C}^3C^b
    \nonumber \\
    &&
    -
    ig\eta\epsilon^{ab}B^3\bar{C}^aC^b
    +
    i\bar{C}^aD^{\xi\mu} D_{\mu} C^a
    \nonumber \\
    &&
    +
    ig\epsilon^{ab}\bar{C}^aD^{\xi\mu}A_\mu^bC^3
    +ig(1-\xi)\epsilon^{ab}\bar C^aA_\mu^b\partial^\mu C^3
    -ig^2\xi\epsilon^{ad}\epsilon^{cb}
    \bar{C}^aC^bA_\mu^cA^{\mu d}
    \nonumber \\
    &&
    +
    g^2\frac{\zeta}{4}\epsilon^{ab}\epsilon^{cd}\bar{C}^a\bar{C}^bC^cC^d
    -
    g^2\eta\bar{C}^3C^3\bar{C}^aC^a
    \biggr\},
\label{eq:S^a}
\end{eqnarray}
where we have defined a covariant derivative $D_\mu^\xi$
in terms of the Abelian gluon $a_\mu$ as
\begin{equation}
D_\mu^\xi{\mit\Phi}^a
  :=(\partial_\mu\delta^{ab}
     -\xi g\epsilon^{ab}a_\mu){\mit\Phi}^b,
\label{eq:Abelian covariant derivative with xi}
\end{equation}
which is abbreviated in the special case of $\xi=1$ as
\begin{equation}
D_\mu{\mit\Phi}^a
 :=D_\mu^{\xi=1}{\mit\Phi}^a
  =(\partial_\mu\delta^{ab}
    -g\epsilon^{ab}a_\mu){\mit\Phi}^b.
\label{eq:Abelian covariant derivative}
\end{equation}

\par
The second functional~$G^{(i)}$ is used to fix the residual
$U(1)$ gauge symmetry.
The possible monomials are of two types:
$\epsilon^{ab3}\bar{C}^3 Y^a Z^b$ and $\delta^{33}\bar{C}^3Y^3$.
Therefore, we obtain
$\epsilon^{ab3}\bar{C}^3 Y^a Z^b
\sim \epsilon^{ab3}\bar{C}^3(\bar{C}^a C^b, A_\mu^a A^\mu{}^b=0)$ 
and 
$
 \delta^{33}\bar{C}^3Y^3
\sim \delta^{33}\bar{C}^3(B^3, \partial^\mu a_\mu) .
$ 
It should be remarked that
the term proportional to $\bar C^3\epsilon^{ab}\bar C^aC^b$
is a candidate for the terms in this functional.
However, such a term has already been included
in Eq.~(\ref{eq:G^a}) as the last term.
Thus the general form of $G^{(i)}$ is given by
\begin{equation}
G^{(i)}
 :=\bar C^3\left[
    \kappa\partial^\mu a_\mu
    +\frac\beta2B^3
   \right].
\label{eq:G^i}
\end{equation}
After performing the BRST transformation, we obtain
\begin{eqnarray}
S^{(i)}
 &:=&
    -i\int d^4x\mbox{\boldmath$\delta$}_{\rm B}G^{(i)}
    \nonumber\\
 &=&\int d^4x\biggl\{
    \frac{\beta}{2}B^3B^3
    +
    \kappa B^3\partial^\mu a_\mu
    +
    i\kappa\bar{C}^3\partial^2C^3
    +
    ig\kappa\bar C^3\epsilon^{ab}\partial^\mu(A_\mu^aC^b)
    \biggr\}.
\end{eqnarray}
\par
The last functional $G^{\rm ex}$ in Eq.~(\ref{eq:S_GF+FP}) includes
exceptional terms.
The possible forms are classified into
$\delta^{ab}X^a Y^b Z^3$, $X^3 Y^3 Z^3$, $\epsilon^{ab}X^a Y^b$ and $X^3$.
The trilinear monomials are 
$
 \delta^{ab}\bar{C}^a Y^b Z^3
\sim \delta^{ab}\bar{C}^a (A_\mu^b a^\mu, \bar{C}^bC^3=0,C^b \bar{C}^3)$
and
$
 \delta^{ab}\bar{C}^3 X^a Y^b
 \sim \bar{C}^3 \delta^{ab}(A_\mu^a A^\mu{}^b, \bar{C}^a C^b) .
$
Moreover, 
$
  X^3 Y^3 Z^3 \sim \bar{C}^3 (a_\mu a^\mu, \bar{C}^3 C^3=0) .
$
The bilinear terms are 
$\epsilon^{ab}X^a Y^b \sim \epsilon^{ab}\bar{C}^a (B^b, \partial^\mu A_\mu^b)$.
The linear term is 
$X^3 \sim \bar{C}^3 (\Lambda^2, \partial^2)$ with a parameter $\Lambda$ of mass dimension one.
Thus $G^{\rm ex}$ is given by
\begin{eqnarray}
G^{\rm ex}
 &:=&g\bar C^3\left[
     \frac\chi2a^\mu a_\mu
     +\frac\varrho2A^{\mu a}A_\mu^a
     +i\varsigma\bar C^aC^a
     \right]
     +g\omega(\Lambda^2+\partial^2)\bar C^3
     \nonumber\\
  & &+\bar C^a\epsilon^{ab}
      (\vartheta\partial^\mu\delta^{bc}
       -\varpi g\epsilon^{bc}a^\mu)A_\mu^c ,
\label{eq:G^ex}
\end{eqnarray}
where we have omitted the bilinear term $\epsilon^{ab}\bar{C}^a B^b$, since it gives a vanishing contribution after the BRST transformation,
$\mbox{\boldmath$\delta$}_{\rm B}(\epsilon^{ab}\bar{C}^a B^b)=0$.
Now we require only the global $U(1)$ gauge symmetry
for the gauge fixing terms
so that the terms included in Eq.~(\ref{eq:G^ex}) are not forbidden in spite of the fact that the diagonal index is not contracted.
The second term in the right hand side of (\ref{eq:G^ex})
becomes a linear term in $B^3$ after carrying out the BRST transformation.
We make use of the dimensional regularization in this paper so that
the divergence coming from the tadpole of $B^3$ does not appear
as a result of perturbative loop expansions.
Therefore we can set the parameter $\omega=0$ without spoiling
the renormalizability.
After performing the BRST transformation, we obtain
\begin{eqnarray}
S^{\rm ex}
 &:=&
    -i\int d^4x\mbox{\boldmath$\delta$}_{\rm B}G^{\rm ex}
    \nonumber\\
 &=&\int d^4x\biggl\{
-
ig\varsigma B^a\bar{C}^3C^a
+
\vartheta\epsilon^{ab}B^aD^\mu A_\mu^b
+
g(\varpi-\vartheta)B^aa_\mu A^{\mu a}
\nonumber \\
&&
+
g\frac{\varrho}{2}B^3A_\mu^a A^{\mu a}
+
g\frac{\chi}{2}B^3a_\mu a^\mu
+
ig\varsigma B^3\bar{C}^aC^a
\nonumber \\
&&
+
i\vartheta\epsilon^{ab}
\bar{C}^aD^\mu D_\mu C^b
+
ig(\varpi-\vartheta)\bar{C}^aa_\mu D^\mu C^a
\nonumber \\
&&
-
ig\vartheta\bar{C}^aD^\mu A_\mu^aC^3
+
ig(\varpi-\vartheta)\bar{C}^a\partial^\mu C^3A_\mu^a
+
ig^2(\varpi-\vartheta)\epsilon^{ab}\bar{C}^aC^3a_\mu A^{\mu b}
\nonumber \\
&&
+ig\varrho\bar C^3A_\mu^aD^\mu C^a
+
ig\chi\epsilon^{ba}\bar{C}^3a_\mu A^{\mu b}C^a
\nonumber \\
&&
+
ig\chi\bar{C}^3a_\mu\partial^\mu C^3
+
ig^2\varpi\epsilon^{bc}\bar{C}^aA_\mu^bC^cA^{\mu a}
-
g^2\varsigma\epsilon^{ab}\bar{C}^3\bar{C}^aC^bC^3
\biggr\}.
\end{eqnarray}

Summing up three functionals and integrating out
the Nakanishi-Lautrup fields, we obtain the most general
form of the gauge fixing term with global $U(1)$ symmetry as
\begin{eqnarray}
S_{\rm GF+FP}
 &=&\int d^4x\bigg\{
+
i\bar{C}^a D^{\mu\xi}D_\mu C^a
+
i\vartheta\epsilon^{ab}\bar{C}^aD^\mu D_\mu C^b
+
i\kappa\bar{C}^3\partial^2C^3
\nonumber \\
&&
+
i\frac{g\kappa}{\beta}(\eta\epsilon^{ab}-\varsigma\delta^{ab})
\bar{C}^aC^b\partial^\mu a_\mu
+
ig(\varpi-\vartheta)\bar{C}^aa_\mu(D^\mu C)^a
\nonumber \\
&&
+
i\frac{g^2\chi}{2\beta}(\eta\epsilon^{ab}-\varsigma\delta^{ab})
\bar{C}^aC^ba_\mu a^\mu
\nonumber \\
&&
+
ig\frac{\alpha-\zeta}{\alpha}\epsilon^{ab}\bar{C}^aD^{\xi\mu}A_\mu^bC^3
+
ig(1-\xi)\epsilon^{ab}\bar{C}^aA_\mu^b\partial^\mu C^3
-
ig\frac{\alpha-\zeta}{\alpha}\vartheta\bar{C}^aD^\mu A^{\mu a}C^3
\nonumber \\
&&
+
ig(\varpi-\vartheta)\bar{C}^a\partial^\mu C^3A_\mu^a
\nonumber \\
&&
+
ig^2\frac{\alpha-\zeta}{\alpha}(\varpi-\vartheta)\epsilon^{ab}
\bar{C}^aC^3a_\mu A^{\mu b}
\nonumber \\
&&
-
ig\kappa\epsilon^{ab}\partial^\mu\bar{C}^3A_\mu^aC^b
+
ig\varrho\bar{C}^3A_\mu^aD^\mu C^a
\nonumber \\
&&
-
i\frac{g}{\alpha}(\eta\epsilon^{ab}-\varsigma\delta^{ab})
\bar{C}^3C^bD^{\xi\mu} A_\mu^a
-
i\frac{g\vartheta}{\alpha}(\eta\delta^{ab}-\varsigma\epsilon^{ba})
\bar{C}^3C^bD^\mu A_\mu^a
\nonumber \\
&&
+
ig\chi\epsilon^{ab}\bar{C}^3a_\mu A^{\mu a}C^b
-
i\frac{g^2}{\alpha}(\varpi-\vartheta)(\eta\epsilon^{ab}-\varsigma\delta^{ab})
\bar{C}^3C^ba_\mu A^{\mu a}
\nonumber \\
&&
+
ig\chi\bar{C}^3a_\mu\partial^\mu C^3
\nonumber \\
&&
+
ig^2\left((-\xi\epsilon^{ad}+\varpi\delta^{ad})\epsilon^{cb}
          +
          \frac{\varrho}{2\beta}(\eta\epsilon^{ab}-\varsigma\delta^{ab})\delta^{cd}
\right)
\bar{C}^aC^bA_\mu^cA^{\mu d}
\nonumber \\
&&
+
g^2\frac{1}{2}\left(\zeta-\frac{\varsigma^2+\eta^2}{\beta}\right)
\delta^{ac}\delta^{bd}
\bar{C}^a\bar{C}^bC^cC^d
\nonumber \\
&&
-
g^2\frac{\alpha-\zeta}{\alpha}(\eta\delta^{ab}-\varsigma\epsilon^{ba})
\bar{C}^3\bar{C}^aC^bC^3
\nonumber \\
&&
-
\frac{1}{2\alpha}
\left(D^{\xi\mu} A_\mu^a
      +
      \vartheta\epsilon^{ab}D^\mu A_\mu^b
      +
      g(\varpi-\vartheta)a_\mu A^{\mu a}\right)^2
\nonumber \\
&&
-
\frac{1}{2\beta}
\left(\kappa\partial^\mu a_\mu
      +
      \frac12g\varrho A_\mu^aA^{\mu a}
      +
      \frac12g\chi a_\mu a^\mu\right)^2
\bigg\}.
\label{eq:the most general action}
\end{eqnarray}
\par
The GF+FP term $S_{\rm GF+FP}$ just obtained has eleven independent
  gauge fixing parameters $\xi$, $\alpha$,
$\zeta$, $\eta$, $\kappa$, $\beta$, $\chi$, $\varrho$, $\varsigma$,
$\vartheta$ and $\varpi$.
If we adopt the most general gauge fixing term, therefore, we must treat
the twelve dimensional parameter space in the Yang-Mills theory with a
gauge coupling constant $g$.
In order to simplify the theory, we try to find the fixed subspace
in which the renormalization group flow is confined
by being protected by some symmetries.
In fact, there are some fixed subspaces in the parameter space
protected by the following symmetries.%
  \footnote{%
    Some of these symmetries were first pointed out
    by Hata and Niigata in Ref.~\cite{HN93}.
  }
\begin{description}
\item[Charge conjugation:]
  The exceptional part~(\ref{eq:G^ex}) breaks
  the ``charge conjugation'' symmetry\cite{HN93}
  under the discrete transformation:
  \begin{equation}
  {\mit\Phi}^1\rightarrow{\mit\Phi}^1,\quad
  {\mit\Phi}^2\rightarrow-{\mit\Phi}^2,\quad
  {\mit\Phi}^3\rightarrow-{\mit\Phi}^3,
  \end{equation}
  where ${\mit\Phi}^A$ denotes all fields.
  Any term belonging to the group~(\ref{eq:2ndgroup}) is not invariant
  under this charge conjugation, while any term belonging to
  the group~(\ref{eq:1stgroup})
  is invariant under the ``charge conjugation''.
  Therefore, by setting the parameter
  $\chi=\varrho=\varsigma=\omega=\vartheta=\varpi=0$,
  the charge conjugation symmetry is recovered.
  However, once we consider the non-perturbative effect, for instance
  ghost--antighost condensation proposed in the previous paper\cite{KS00a},
  the ``charge conjugation'' invariance%
  \footnote{%
    There are two types ghost--antighost condensation,
    $C^a\bar C^a=C^1\bar C^1+C^2\bar C^2$ and
    $\epsilon^{ab}\bar C^aC^b=C^1\bar C^2-C^2\bar C^1$.
    The ``charge conjugation'' invariance is broken by the latter one.
    (See Ref.~\cite{KS00a} for more details.) 
  }
  is not expected to hold.
\item
  [Translational invariance for $\bar C^3$:]
  By setting the parameter to
  $\eta=0$ and $\chi=\varrho=\varsigma=\omega=0$,
  the GF+FP term respects a global symmetry under the translation
  of the diagonal antighost $\bar C^3(x)$ as
  $\bar C^3(x)\rightarrow\bar C^3(x)+\bar\theta^3$
  where $\bar\theta^3$ is a constant Grassmann variable.
  This is because the diagonal antighost $\bar C^3$
  appears only in the differentiated form  $\partial_\mu\bar C^3$
  for this choice of the parameters.
  Then the translational symmetry of $\bar C^3$ exists in the theory.

\item
  [Translational invariance for $C^3$:]
  By setting the parameter to $\alpha=\zeta$,
  the action has a global symmetry
  under the translation of the diagonal ghost as
  $C^3(x)\rightarrow C^3(x)+\theta^3$
  where $\theta^3$ is a constant Grassmann variable.
  In the similar manner to the previous case,
  we can confirm that the translational
  symmetry of $C^3$ exists in this case.

\item
  [Implicit residual $U(1)$ invariance:]
  By setting the parameter to
  $\xi=1$, $\chi=\varrho=\varsigma=\omega=0$ and $\varpi=\vartheta$,
  the action has the residual $U(1)$ gauge symmetry mentioned in the previous
  paper\cite{KS00b}, although the gauge fixing
  for the residual $U(1)$ gauge symmetry has already been accomplished.
  As we have mentioned in the previous paper\cite{KS00b},
  there is the $U(1)$ gauge symmetry if the diagonal gluon does not
  appear in the action after replacing
  all the derivatives with the Abelian covariant derivative defined by
  (\ref{eq:Abelian covariant derivative})
  except for a quadratic term as $(\partial^\mu a_\mu)^2$.
  In the view of the background field method\cite{Abbott82}, there is
  a gauge symmetry with respect to the background diagonal field.
\begin{eqnarray}
S_{\rm GF+FP}
 &=&\int d^4x\bigg\{
-
\frac{1+\vartheta^2}{2\alpha}
\left(D^{\mu} A_\mu^a\right)^2
-
\frac{1}{2\beta}
\left(\kappa\partial^\mu a_\mu\right)^2
\nonumber \\
&&
+
i\bar{C}^a D^{\mu}D_\mu C^a
+
i\vartheta\epsilon^{ab}\bar{C}^aD^\mu D_\mu C^b
+
i\kappa\bar{C}^3\partial^2C^3
\nonumber \\
&&
+
i\frac{g}{\alpha}
\left(\epsilon^{ab}-\vartheta\delta^{ab}\right)
\left[(\alpha-\zeta)\bar{C}^aD^{\mu}A_\mu^bC^3
      +\eta\bar{C}^3C^aD^{\mu} A_\mu^b\right]
\nonumber \\
&&
+
i\frac{g\kappa}{\beta}\eta\epsilon^{ab}
\bar{C}^aC^b\partial^\mu a_\mu
-
ig\kappa\epsilon^{ab}\partial^\mu\bar{C}^3A_\mu^aC^b
%
%
-
ig^2\epsilon^{ad}\epsilon^{cb}
\bar{C}^aC^bA_\mu^cA^{\mu d}
\nonumber \\
&&
+
\frac{g^2}{2}\left(\zeta-\frac{\eta^2}{\beta}\right)
\bar{C}^a\bar{C}^bC^aC^b
-
g^2\frac{\alpha-\zeta}{\alpha}\eta
\bar{C}^3\bar{C}^aC^aC^3
\bigg\}.
\label{eq:residual U(1)}
\end{eqnarray}


\item
  [Anti-BRST symmetry:]
  By setting the parameter to
  $\chi=\varrho=\varsigma=\omega=0$, $\vartheta=\varpi=0$,
  $1-\xi-\kappa=0$ and
  $\alpha-\beta+\eta-\zeta=0$,
  the action has the anti-BRST invariance.
  Then the action is given by
\begin{eqnarray}
S_{\rm GF}
 &\equiv&
   \int d^4x\biggl\{
   i\mbox{\boldmath$\delta$}_{\rm B}
    \bar{\mbox{\boldmath$\delta$}}_{\rm B}
    \biggl[\frac12A^{\mu a}A_\mu^a
           -\frac i2(\eta+\zeta)C^a\bar C^a
           +\frac\kappa2a^\mu a_\mu
           -\eta iC^3\bar C^3
           \biggr]
    \nonumber\\
 & &+\frac12(\alpha-\eta-\zeta)B^AB^A
    \biggr\}.
\label{eq:BRST--Anti-BRST}
\end{eqnarray}
Here, the second term in the integrand of the right hand side of the
Eq.~(\ref{eq:BRST--Anti-BRST}) is not exact in the combined BRST and
anti-BRST transformations,
$
\mbox{\boldmath$\delta$}_{\rm B}
\bar{\mbox{\boldmath$\delta$}}_{\rm B}
$,
differently from the first term.
However, the second term is both BRST and anti-BRST invariant since
$B^AB^A
  =-i\mbox{\boldmath$\delta$}_{\rm B}
    (\bar C^AB^A)
  =i\bar{\mbox{\boldmath$\delta$}}_{\rm B}
    (C^AB^A)
$.

\item
  [Global $SU(2)$ invariance:]
  After setting the parameter to
  $\chi=\varrho=\varsigma=\omega=0$, $\vartheta=\varpi=0$,
  $\xi=0$, $\kappa=1$, $\alpha=\beta$ and $\zeta=\eta$,
  the action has the global $SU(2)$ invariance.
  Then the action is given by
\begin{equation}
S_{\rm GF}
 \equiv
  -\int d^4x
   i\mbox{\boldmath$\delta$}_{\rm B}
    \biggl\{\bar C^A
    \biggl[\partial^\mu A_\mu^A
           +\frac\alpha2B^A
           -\frac\zeta2\epsilon^{ABC}\bar C^BC^C
           \biggr]
    \biggr\}.
\label{eq:SU2 action}
\end{equation}
It is easy to see that this choice of parameters is
a spacial case of the anti-BRST symmetric case.
As a result, the action~(\ref{eq:SU2 action}) is recast into 
\begin{equation}
S_{\rm GF}
 =\int d^4x
  \left[
   i\mbox{\boldmath$\delta$}_{\rm B}
    \bar{\mbox{\boldmath$\delta$}}_{\rm B}
    \left(\frac12{\cal A}_\mu^A{\cal A}^{\mu A}
          -\frac\xi2i{\cal C}^A\bar{\cal C}^A\right)
    +\frac{\xi^\prime}2{\cal B}^A{\cal B}^A
    \right],
\end{equation}
by introducing $\xi$ and $\xi^\prime$ as $\alpha=\xi+\xi^\prime$,
$\zeta=\xi/2$.
This form agrees with the global $SU(2)$ invariant action which is invariant under the BRST and anti-BRST transformation obtained by Baulieu and Thierry-Mieg\cite{BT82}.

\item
  [FP conjugation invariance:]
  After setting the parameter to
  $\chi=\varrho=\varsigma=\omega=0$, $\vartheta=\varpi=0$,
  $\alpha=\zeta+\eta$, $1-\xi-\kappa=0$ and $\kappa(\beta-2\eta)=0$
  and integrating out $B^3$ and $B^a$,
the action has the invariance under the FP ghost conjugation:
\begin{equation}
C^A\rightarrow\pm\bar C^A,
\quad
\bar C^A\rightarrow\mp C^A,
\quad
{\cal A}_\mu^A\rightarrow{\cal A}_\mu^A.
\end{equation}
In the case of $\kappa\ne0$, i.e., $\beta=2\eta$,
the anti-BRST symmetry is also recovered and we obtain 
\begin{equation}
S_{\rm GF}
  =\int d^4x\biggl\{
   i\mbox{\boldmath$\delta$}_{\rm B}
    \bar{\mbox{\boldmath$\delta$}}_{\rm B}
    \biggl[\frac12A^{\mu a}A_\mu^a
           -\frac i2\alpha C^a\bar C^a
           +\frac\kappa2a^\mu a_\mu
           -\frac i2\beta C^3\bar C^3
           \biggr]
   \biggl\}
\label{eq:SL(2,R)}.
\end{equation}
It is identical to the BRST--anti-BRST exact part
of the FP+GF term~(\ref{eq:BRST--Anti-BRST}) previously discussed.
Another case, $\kappa=0$, is very delicate
since the gauge fixing term of the Abelian gluon is eliminated
by the naive limit of $\kappa\to0$.

\item
  [$SL(2,R)$ symmetry:]
  After setting the parameter to
  $\chi=\varrho=\varsigma=\omega=0$, $\vartheta=\varpi=0$,
  $\alpha=\zeta+\eta$, $1-\xi-\kappa=0$ and $\kappa(\beta-2\eta)=0$
  and integrating out $B^3$ and $B^a$,
  just as in the FP conjugation invariance,
  the GF+FP term also become invariant under the two transformations:
\begin{equation}
\delta_+\bar C^A(x)=C^A(x),
\qquad
\delta_+(\mbox{other fields})=0,
\end{equation}
\begin{equation}
\delta_-C^A(x)=\bar C^A(x),
\qquad
\delta_-(\mbox{other fields})=0.
\end{equation}
These symmetries are $SL(2,R)$ symmetry for the multiplet of ghost and
antighost $(C,\bar C)$, see e.g. Ref.~\cite{KS00a}.

\end{description}


When we wish to perform thorough analyses for phenomena
with unbroken global $U(1)$ symmetry,
it is desirable to employ
the most general action~(\ref{eq:the most general action}).
However, it is very tedious work
so that we should require some additional symmetries
and restrict the gauge parameter space properly.

Now we define the most general MA gauge.
In the analysis of the conventional MA gauge,
the residual $U(1)$ symmetry is the most important symmetry
since the MA gauge condition is originally defined as follows.
The MA gauge is obtained by minimizing the functional $R[A^U]$
with respect to the local gauge transformation $U(x)$ of $A_\mu^a(x)$.
Here, $R[A]$ is defined as the functional of off-diagonal gluon fields,
\begin{equation}
R[A] := \int d^4x {1 \over 2} A_\mu^a(x) A^\mu{}^a(x) .
\end{equation}
Then we obtain the differential form of the MA gauge condition,
\begin{equation}
D_\mu A^\mu{}^a
 \equiv\left(\partial_\mu\delta^{ab} -g\epsilon^{ab}a_\mu\right) A^\mu{}^b
 =0.
\label{eq:MA gauge condition}
\end{equation}
To adopt this condition~(\ref{eq:MA gauge condition}),
we must set $\xi=1$, $\chi=\varrho=\varsigma=\omega$ and $\varpi=\vartheta$
as we mentioned in Implicit residual $U(1)$ invariance
and (\ref{eq:residual U(1)}).

A remarkable difference between our gauge fixing procedure 
and that of the previous works (Min, Lee and Pac \cite{MLP85} and Hata and Niigata\cite{HN93})
is the existence of a new parameter $\kappa$. 
If we do not require the recovery of global $SU(2)$ gauge symmetry
in our gauge fixing,  then there is no need to set the parameter
$\kappa$ to 1 against Ref.~\cite{HN93}.
At first sight, the parameter $\kappa$ can be absorbed by rescaling
the diagonal ghost $C^3$ and diagonal antighost $\bar C^3$.
However, such a rescaling is not always valid.
For instance, requiring the residual $U(1)$ gauge symmetry to
the gauge fixing term~(\ref{eq:SL(2,R)}) which has
invariance under $SL(2,R)$ symmetry and FP conjugation,
we must set $\kappa=0$ since there is a relation $1-\xi-\kappa=0$.
Therefore, in this case, we cannot absorb the parameter $\kappa$
by rescaling the field.
On the contrary, setting $\kappa=1$
in the gauge fixing term~(\ref{eq:SL(2,R)}),
the gauge fixing condition becomes $\partial_\mu A^{\mu a}=0$
which is identical to the ordinary Lorentz gauge condition
so that it is not the MA gauge in the ordinary sense
in spite of the breaking of the global $SU(2)$ symmetry
for $\alpha\ne\beta$.

Another advantage of introducing the parameter $\kappa$
is that the symmetry of the {\it renormalized} theory under the FP conjugation
(i.e., the symmetry of the renormalized theory
 under the exchange of the ghost and antighost)
is easily examined for the renormalized theory,
since the renormalized ghost and antighost fields are defined
through the same renormalization factor.

\section{The minimum choice of the gauge fixing terms} %
By requiring some of the symmetries listed in the previous section,
that is, charge conjugation,
translational invariance of the diagonal ghost $C^3$
or antighost $\bar C^3$ and implicit $U(1)$ gauge symmetry,
we obtain the minimum choice of the renormalizable MA gauge.
Setting parameters as
\begin{equation}
\alpha=\zeta,\quad
\xi=1,\quad
\mbox{and}\quad
\eta=\chi=\varrho=\varsigma=\omega=\vartheta=\varpi=0,
\end{equation}
we arrive at the gauge fixing terms with three parameters
$\alpha$, $\beta$ and $\kappa$.
\begin{eqnarray}
S_{\rm GF}
 &:=&
    i\int d^4x
     \mbox{\boldmath$\delta$}_{\rm B}
     \bar{\mbox{\boldmath$\delta$}}_{\rm B}
     \left[\frac12A^{\mu a}A_\mu^a
           -\frac\alpha2iC^a\bar C^a\right]
    -i\int d^4x
     \mbox{\boldmath$\delta$}_{\rm B}
     \left\{
     \bar C^3\left[
     \kappa\partial^\mu a_\mu
     +\frac\beta2B^3
    \right]
    \right\}
    \nonumber\\
 &=&\int d^4x
    \biggl\{
    B^aD^\mu A_\mu^b
    +\frac\alpha2B^aB^a
    +i\bar C^aD^2C^b
    -ig^2\epsilon^{ad}\epsilon^{cb}\bar C^aC^bA^{\mu c}A_\mu^d
    \nonumber\\
 & &+ig\epsilon^{ab}\bar C^a(D^\mu A_\mu^b)C^3
    -i\alpha g\epsilon^{ab}B^a\bar C^bC^3
    +\frac\alpha4g^2\epsilon^{ab}\epsilon^{cd}\bar C^a\bar C^bC^cC^d
    \nonumber\\
 & &+\kappa B^3\partial^\mu a_\mu
    +\frac\beta2B^3B^3
    +i\kappa\bar C^3\partial^2C^3
    +i\kappa\bar C^3\partial^\mu(g\epsilon^{bc}A_\mu^bC^c)
    \biggr\}.
\label{eq:S_kappa1}
\end{eqnarray}
By integrating out $B^3$ and $B^a$, we obtain
\begin{eqnarray}
S_{\rm GF}
 &=&\int d^4x
    \biggl\{
    -\frac1{2\alpha}(D^\mu A_\mu^a)^2
    -\frac{\kappa^2}{2\beta}(\partial^\mu a_\mu)^2
    \nonumber\\
 & &+i\bar C^aD^2C^b
    -ig^2\epsilon^{ad}\epsilon^{cb}\bar C^aC^bA^{\mu c}A_\mu^d
    +\frac\alpha4g^2\epsilon^{ab}\epsilon^{cd}\bar C^a\bar C^bC^cC^d
    \nonumber\\
 & &+i\kappa\bar C^3\partial^2C^3
    +i\kappa\bar C^3\partial^\mu(g\epsilon^{bc}A_\mu^bC^c)
    \biggr\}.
\label{eq:S_kappa2}
\end{eqnarray}

In the following subsections,
we consider the renormalizability of this action
in the two different schemes.
In the Scheme~I, the parameter $\kappa$ is absorbed
by rescaling the antighost field.
In this case, we must distinguish the renormalization factors
of the diagonal ghost and diagonal antighost fields.  
This approach is valid except for the case of $\kappa=0$,
in which the rescaling of the antighost field is ill-defined.
In the Scheme~$\rm I\!I$, the parameter $\kappa$ is
left explicitly.
In this case we can equate the renormalization factors
of the diagonal ghost and diagonal antighost fields.
This approach is valid even if $\kappa=0$,
and we demonstrate that the case of $\kappa=0$ is meaningful
from the viewpoint of renormalizability and symmetries.

\subsection{Scheme I}
In the case of $\kappa\ne0$,
the parameter $\kappa$ can be absorbed by rescaling the diagonal antighost
field and parameter $\beta$ as
\begin{equation}
\bar C^3\rightarrow\frac{\bar C^3}\kappa,
\quad
\beta\rightarrow\kappa^2\beta,
\label{eq:absorption of kappa}
\end{equation}
and we obtain
\begin{eqnarray}
S_{\rm GF}
 &=&\int d^4x
    \biggl\{
    -\frac1{2\alpha}(D^\mu A_\mu^a)^2
    -\frac1{2\beta}(\partial^\mu a_\mu)^2
    \nonumber\\
 & &+i\bar C^aD^2C^b
    -ig^2\epsilon^{ad}\epsilon^{cb}\bar C^aC^bA^{\mu c}A_\mu^d
    +\frac\alpha4g^2\epsilon^{ab}\epsilon^{cd}\bar C^a\bar C^bC^cC^d
    \nonumber\\
 & &+i\bar C^3\partial^2C^3
    +i\bar C^3\partial^\mu(g\epsilon^{bc}A_\mu^bC^c)
    \biggr\}.
\label{eq:S_kappa3}
\end{eqnarray}
We notice that the diagonal ghost $C^3$ does not appear
in the interaction terms in (\ref{eq:S_kappa3}).
Therefore {\it we need not take account of the internal diagonal ghost
in the calculation of perturbative loop expansions}.
The beta-function,
the anomalous dimensions of the diagonal gluon $a_\mu$
and off-diagonal gluon $A_\mu^a$,
and RG functions of two gauge fixing parameters
$\alpha$ and $\beta$ have already been obtained
in previous papers\cite{KS00b,Shinohara01a}.
In this paper, we determine the anomalous dimension
of the ghost field $C$ and antighost field $\bar C$ by making use of
the dimensional regularization at the one-loop level.

\subsubsection{Feynman rules} %
From the total action:
\begin{equation}
S:=S_{\rm YM}
   +S_{\rm GF},
\label{eq:S}
\end{equation}
with the Yang-Mills action
\begin{equation}
S_{\rm YM}
 =-\int d^4x\frac14F_{\mu\nu}^AF^{\mu\nu A},
\label{eq:S_YM}
\end{equation}
we obtain the following Feynman rules.


\unitlength=0.001in
\begin{figure}[tbp]
\begin{center}
\begin{picture}(5600,2400)
\put(0,2200){\mbox{(a)}}%
\put(550,2100){%
   \put(-150,0){\epsfysize=5mm \epsfbox{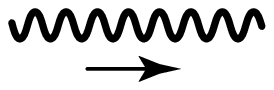}}%
   \put(-300,110){\mbox{$\mu$}}%
   \put(550,110){\mbox{$\nu$}}%
   \put(80,-60){\mbox{$p$}}%
   }%
\put(1400,2200){\mbox{(b)}}%
\put(1950,2100){%
   \put(0,0){\epsfysize=5mm \epsfbox{apropa.eps}}%
   \put(-280,110){\mbox{$a,\mu$}}%
   \put(700,110){\mbox{$b,\nu$}}%
   \put(230,-50){\mbox{$p$}}%
   }%
\put(3050,2200){\mbox{(c)}}%
\put(3450,2200){%
   \put(-50,0){\epsfysize=2mm \epsfbox{ghost.eps}}%
   \put(200,-100){\mbox{$p$}}%
   }%
\put(4350,2200){\mbox{(d)}}%
\put(4800,2200){%
   \put(-50,0){\epsfysize=2mm \epsfbox{ghost.eps}}%
   \put(-180,0){\mbox{$a$}}%
   \put(630,0){\mbox{$b$}}%
   \put(200,-100){\mbox{$p$}}%
   }%
\put(0,1700){\mbox{(e)}}%
\put(200,1000){%
   \put(0,0){\epsfysize=20mm \epsfbox{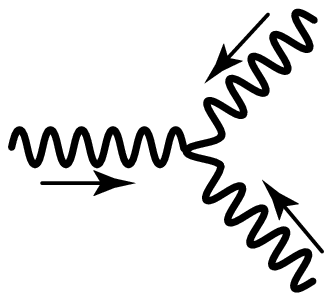}}%
   \put(150,200){\mbox{$p$}}%
   \put(580,750){\mbox{$q$}}%
   \put(880,150){\mbox{$r$}}%
   \put(100,500){\mbox{$3,\mu$}}%
   \put(860,600){\mbox{$a,\rho$}}%
   \put(470,30){\mbox{$b,\sigma$}}%
   }%
\put(1450,1700){\mbox{(f)}}%
\put(1700,1000){%
   \put(0,0){\epsfysize=20mm \epsfbox{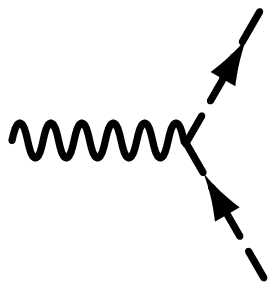}}%
   \put(570,700){\mbox{$p$}}%
   \put(700,150){\mbox{$q$}}%
   \put(100,500){\mbox{$3,\mu$}}%
   \put(700,600){\mbox{$a$}}%
   \put(570,30){\mbox{$b$}}%
   }%
\put(2700,1700){\mbox{(g)}}%
\put(3000,1000){%
   \put(0,0){\epsfysize=20mm \epsfbox{acc.eps}}%
   \put(570,700){\mbox{$p$}}%
   \put(700,150){\mbox{$q$}}%
   \put(100,500){\mbox{$c,\mu$}}%
   \put(720,600){\mbox{$3$}}%
   \put(570,30){\mbox{$b$}}%
   }%
\put(4020,1700){\mbox{(h)}}%
\put(4500,1000){%
   \put(0,0){\epsfysize=20mm \epsfbox{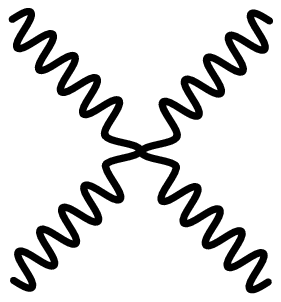}}%
   \put(-250,600){\mbox{$3,\mu$}}%
   \put(-250,100){\mbox{$3,\nu$}}%
   \put(730,600){\mbox{$a,\rho$}}%
   \put(730,100){\mbox{$b,\sigma$}}%
   }%
\put(0,700){\mbox{(i)}}%
\put(400,0){%
   \put(0,0){\epsfysize=20mm \epsfbox{aaaa.eps}}%
   \put(-250,600){\mbox{$a,\mu$}}%
   \put(-250,100){\mbox{$b,\nu$}}%
   \put(730,600){\mbox{$c,\rho$}}%
   \put(730,100){\mbox{$d,\sigma$}}%
   }%
\put(1500,700){\mbox{(j)}}%
\put(1950,0){%
   \put(0,0){\epsfysize=20mm \epsfbox{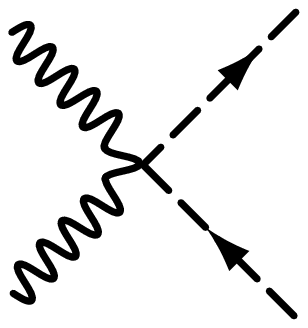}}%
   \put(-250,600){\mbox{$3,\mu$}}%
   \put(-250,100){\mbox{$3,\nu$}}%
   \put(650,600){\mbox{$a$}}%
   \put(650,100){\mbox{$b$}}%
   }%
\put(2850,700){\mbox{(k)}}%
\put(3300,0){%
   \put(0,0){\epsfysize=20mm \epsfbox{aacc.eps}}%
   \put(-250,600){\mbox{$c,\mu$}}%
   \put(-250,100){\mbox{$d,\nu$}}%
   \put(650,600){\mbox{$a$}}%
   \put(650,100){\mbox{$b$}}%
   }%
\put(4300,700){\mbox{(l)}}%
\put(4550,0){%
   \put(0,0){\epsfysize=20mm \epsfbox{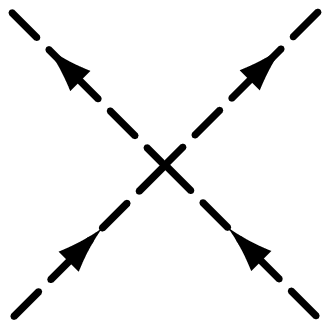}}%
   \put(0,580){\mbox{$a$}}%
   \put(0,130){\mbox{$c$}}%
   \put(750,600){\mbox{$b$}}%
   \put(750,100){\mbox{$d$}}%
   }%
\end{picture}
\caption[]{%
    The wavy line corresponds to the gluon,
    and the broken line corresponds to the ghost or antighost.
    The graphs in (a) and (b) represent respectively diagonal and
    off-diagonal gluons while the graphs in (c) and (d) represent
    respectively diagonal and off-diagonal ghosts.
    The graphs in (e), (f) and (g) represent the three-point vertex
    and the graphs in (h), (i), (j), (k) and (l)
    represent the four-point vertex.
}
\label{fig:Feynman Rules}
\end{center}
\end{figure}

\paragraph{Propagators} %
\begin{enumerate}
\item[(a)] diagonal gluon propagator:
\begin{equation}
iD_{\mu\nu}
  =-\frac i{p^2}
    \left[g_{\mu\nu}
          -\left(1-\beta\right)
           \frac{p_\mu p_\nu}{p^2}\right].
\label{eq:propagator of diagonal gluon}
\end{equation}

\item[(b)] off-diagonal gluon propagator:
\begin{equation}
iD_{\mu\nu}^{ab}
  =-\frac i{p^2}
    \left[g_{\mu\nu}-(1-\alpha)\frac{p_\mu p_\nu}{p^2}\right]\delta^{ab} .
\label{eq:propagator of off-diagonal gluon}
\end{equation}

\item[(c)] diagonal ghost propagator:
\begin{equation}
i\Delta
  =-\frac1{p^2}.
\end{equation}

\item[(d)] off-diagonal ghost propagator:
\begin{equation}
i\Delta^{ab}
  =-\frac1{p^2}\delta^{ab}.
\end{equation}
\end{enumerate}

\paragraph{Three-point vertices} %
\begin{enumerate}
\item[(e)] One diagonal and two off-diagonal gluons:
\begin{eqnarray}
&&i\left<a_\mu(p)A_\rho^a(q)A_\sigma^b(r)\right>_{\rm bare}
  \nonumber\\
&&\textstyle
 =g\epsilon^{ab}
  \left[(q-r)_\mu g_{\rho\sigma}
        +\left\{r-p+\frac q\alpha\right\}_\rho g_{\sigma\mu}
        +\left\{p-q-\frac r\alpha\right\}_\sigma g_{\mu\rho}
  \right].
\end{eqnarray}

\item[(f)]
One diagonal gluon,
one off-diagonal ghost and one off-diagonal antighost:
\begin{equation}
i\left<\bar C^a(p)C^b(q)a_\mu\right>_{\rm bare}
  =-i(p+q)_\mu g\epsilon^{ab}.
\end{equation}

\item[(g)]
One off-diagonal gluon,
one off-diagonal ghost and one diagonal antighost:
\begin{equation}
i\left<\bar C^3(p)C^b(q)A_\mu^c\right>_{\rm bare}
  =-ig\epsilon^{cb}p_\mu.
\end{equation}


\end{enumerate}

\paragraph{Four-point vertices} %
\begin{enumerate}
\item[(h)] Two diagonal gluons and two off-diagonal gluons:
\begin{equation}
i\left<A_\mu^3A_\nu^3
       A_\rho^aA_\sigma^b\right>_{\rm bare}
  =-ig^2\delta^{ab}
    \left[2g_{\mu\nu}g_{\rho\sigma}
          -\left(1-\frac1\alpha\right)
          (g_{\mu\rho}g_{\nu\sigma}+g_{\mu\sigma}g_{\nu\rho})\right].
\end{equation}

\item[(i)] Four off-diagonal gluons:
\begin{equation}
  i\left<A_\mu^aA_\nu^bA_\rho^cA_\sigma^d\right>_{\rm bare}
    =-i2g^2\bigl[\epsilon^{ab}\epsilon^{cd}I_{\mu\nu,\rho\sigma}
            +\epsilon^{ac}\epsilon^{bd}I_{\mu\rho,\nu\sigma}
            +\epsilon^{ad}\epsilon^{bc}I_{\mu\sigma,\nu\rho}],
\end{equation}
where
$I_{\mu\nu,\rho\sigma}
  :=(g_{\mu\rho}g_{\nu\sigma}
     -g_{\mu\sigma}g_{\nu\rho})/2
$
.

\item[(j)]
Two diagonal gluons,
one off-diagonal ghost and one off-diagonal antighost:
\begin{equation}
i\left<\bar C^aC^bA_\mu^3A_\nu^3\right>_{\rm bare}
  =2g^2\delta^{ab}g_{\mu\nu}.
\end{equation}

\item[(k)]
Two off-diagonal gluons,
one off-diagonal ghost and one off-diagonal antighost:
\begin{equation}
i\left<\bar C^aC^bA_\mu^cA_\nu^d\right>_{\rm bare}
  =g^2\left[\epsilon^{ad}\epsilon^{cb}
            +\epsilon^{ac}\epsilon^{db}\right]g_{\mu\nu}.
\end{equation}

\item[(l)]
Two off-diagonal ghosts and two off-diagonal antighosts:
\begin{equation}
i\left<\bar C^a\bar C^bC^cC^d\right>_{\rm bare}
  =ig^2\alpha\epsilon^{ab}\epsilon^{cd}.
\end{equation}

\end{enumerate}

\subsubsection{Counterterms} %
In order to construct the renormalized theory, we define the following
renormalized fields%
\footnote{%
Here, the renormalization factors for
off-diagonal ghost $C^a$ and off-diagonal antighost $\bar C^a$
can be identical to each other.
On the other hand, we must introduce different renormalization factors
$Z_c$ and $Z_{\bar c}$
for diagonal ghost $C^3$ and diagonal antighost $\bar C^3$ respectively,
since we absorbed the parameter $\kappa$
to the diagonal antighost field $\bar C^3$.
See Appendix, for more details.}
and parameters:
\begin{equation}
\begin{array}{c}
\begin{array}{ccc}
a_\mu=Z_a^{1/2}a_{{\rm R}\mu}, &
C^3=Z_c^{1/2}C_{{\rm R}}^3, &
\bar C^3=Z_{\bar c}^{1/2}\bar C_{{\rm R}}^3, \\[2mm]
A_\mu^a=Z_A^{1/2}A_{{\rm R}\mu}^a, &
C^a=Z_C^{1/2}C_{{\rm R}}^a, &
\bar C^a=Z_C^{1/2}\bar C_{{\rm R}}^a,
\end{array}\\[6mm]
g=Z_gg_{\rm R},
\quad
\alpha=Z_\alpha\alpha_{\rm R},
\quad
\beta=Z_{\beta}\beta_{\rm R}.
\end{array}
\label{eq:renormalization}
\end{equation}
By substituting
the above renormalization relations~(\ref{eq:renormalization})
into the action~(\ref{eq:S}), we obtain
\begin{equation}
S=S_{\rm R}
  +\Delta S_{\rm gauge}
  +\Delta S_{\rm ghost}.
\end{equation}
Here $S_{\rm R}$ is the renormalized action obtained
from the bare action~(\ref{eq:S})
by replacing all the fields and parameters with the renormalized ones,
while $\Delta S_{\rm ghost}$ and $\Delta S_{\rm gauge}$ are
counterterms with and without ghost fields respectively.
In this paper we focus on the renormalizability of the terms
with ghost fields $\Delta S_{\rm ghost}$.
This is explicitly given by
\begin{eqnarray}
\Delta S_{\rm ghost}
 &=&\int d^4x\biggl\{
    i\delta_a
     \bar C_{\rm R}^aD_{\rm R}^2C_{\rm R}^a
    +i\delta_b
     \bar C_{\rm R}^3\partial^2C_{\rm R}^3
    \nonumber\\
 & &-i\delta_c
     g_{\rm R}^2\epsilon^{ad}\epsilon^{cb}
     \bar C_{\rm R}^aC_{\rm R}^bA_{\rm R}^{\mu c}A_{{\rm R}\mu}^d
    +\delta_d
     \frac{\alpha_{\rm R}}4g_{\rm R}^2\epsilon^{ab}\epsilon^{cd}
     \bar C_{\rm R}^a\bar C_{\rm R}^bC_{\rm R}^cC_{\rm R}^d
    \nonumber\\
 & &+i\delta_e
     \bar C_{\rm R}^3\partial^\mu
     (g_{\rm R}\epsilon^{bc}A_{{\rm R}\mu}^bC_{\rm R}^c)
    \biggr\},
\label{eq:dS_ghost}
\end{eqnarray}
where we have defined the renormalized Abelian covariant derivative
$D_{\rm R}$ by
\begin{equation}
D_{{\rm R}\mu}{\mit\Phi}^a
  :=\left(\partial_\mu\delta^{ab}
         -g_{\rm R}\epsilon^{ab}a_{{\rm R}\mu}\right){\mit\Phi}^b.
\end{equation}
Abelian covariant derivative itself does not change
under the renormalization.
Indeed, substituting the renormalized relations~(\ref{eq:renormalization})
into the definition of the bare Abelian covariant
derivative~(\ref{eq:Abelian covariant derivative}) and using the
relation $Z_g=Z_a^{-1/2}$ due to the implicit residual $U(1)$ gauge
symmetry pointed out in the previous paper\cite{KS00b},
we obtain
\begin{eqnarray}
D_\mu{\mit\Phi}^a
 &=&\left(\partial_\mu\delta^{ab}
          -g\epsilon^{ab}a_\mu\right){\mit\Phi}^b
    \nonumber\\
 &=&\left(\partial_\mu\delta^{ab}
          -Z_gZ_a^{1/2}g_{\rm R}\epsilon^{ab}a_{{\rm R}\mu}\right)
    {\mit\Phi}^b
    \nonumber\\
 &=&\left(\partial_\mu\delta^{ab}
          -g_{\rm R}\epsilon^{ab}a_{{\rm R}\mu}\right){\mit\Phi}^b
    \nonumber\\
 &=&D_{{\rm R}\mu}{\mit\Phi}^a.
\end{eqnarray}

The coefficients
$\delta=(\delta_a,\delta_b,\delta_c,\delta_d,\delta_e)$
in the counter terms~(\ref{eq:dS_ghost}) are related
to the renormalization factors~$Z_X=(Z_c, Z_{\bar c}, Z_C)$ as
\begin{equation}
\begin{array}{ll}
\delta_a
 =Z_C-1,&
\delta_b
 =Z_c^{1/2}Z_{\bar c}^{1/2}-1,\\[2mm]
\delta_c
 =Z_g^2Z_CZ_A-1,&
\delta_d
 =Z_\alpha Z_g^2Z_C^2-1,\\[2mm]
\delta_e
 =
  Z_{\bar c}^{1/2}
  Z_g
  Z_A^{1/2}Z_C^{1/2}-1.&
\label{eq:deltas}
\end{array}
\end{equation}
Therefore we can determine the renormalization factors $Z$s
by calculating $\delta$s.

\subsubsection{Anomalous dimensions and RG functions} %
In this subsection, we determine the renormalization factors and
anomalous dimensions of the fields and parameters.
The renormalization factor $Z_X$ is expanded
order by order of the loop expansion as
\begin{equation}
Z_X=1+Z_X^{(1)}+Z_X^{(2)}+\cdots,
\end{equation}
where $Z_X^{(n)}$ is the $n$th order contribution.
The anomalous dimension of the respective field $X=Z_X^{1/2}X_{\rm R}$
is defined by
\begin{equation}
\gamma_X
 :=\frac12\mu\frac{\partial}{\partial\mu}\ln Z_X
 :=\frac12\mu\frac{\partial}{\partial\mu}Z_X^{(1)}+\cdots,
\label{eq:def. of anomalous dimension 1}
\end{equation}
and the RG function of the respective parameter
$Y=Z_YY_{\rm R}$ is defined by
\begin{equation}
\gamma_Y
 :=\mu\frac{\partial}{\partial\mu}Y_{\rm R}
 :=-Y_{\rm R}\mu\frac{\partial}{\partial\mu}Z_Y^{(1)}+\cdots.
\label{eq:def. of anomalous dimension 2}
\end{equation}

The anomalous dimension of the diagonal gluon $a_\mu$
can be determined by requiring the renormalizability for the transverse
part of the diagonal gluon propagator.
On the other hand, the RG function of the Abelian gauge
fixing parameter $\beta$ can be determined by requiring the
renormalizability for the longitudinal part of
the diagonal gluon propagator.
Similarly, the anomalous dimensions of the off-diagonal gluons $A_\mu^a$
and the RG function of the gauge fixing parameter $\alpha$
can be respectively determined
by considering the transverse and longitudinal part of
of off-diagonal gluon propagators.
Then the renormalization factors $Z_a$, $Z_{\beta}$, $Z_A$ and $Z_\alpha$
are obtained by calculating the counterterms $\Delta S_{\rm gauge}$.
Moreover, from the counterterms $\Delta S_{\rm gauge}$
we can calculate also the RG function of the QCD coupling
constant $g$, that is, the $\beta$-function.
These renormalization factors have already been calculated
in Ref.~\cite{KS00b,Shinohara01a}
by making use of the dimensional regularization.
Consequently, the renormalization factors are given as
\begin{equation}
Z_a^{(1)}
 =Z_{\beta}^{(1)}
 =\frac{22}3\frac{(\mu^{-\epsilon}g_{\rm R})^2}{(4\pi)^2\epsilon},
\label{eq:Z_a}
\end{equation}
\begin{equation}
Z_A^{(1)}
 = \frac{(g_{\rm R}\mu^{-\epsilon})^2}{(4\pi)^2\epsilon}
   \left[\frac{17}6-\frac{\alpha_{\rm R}}2-\beta_{\rm R}
   \right],
\label{eq:Z_A}
\end{equation}
\begin{equation}
Z_\alpha^{(1)}
 =\frac{(g_{\rm R}\mu^{-\epsilon})^2}{(4\pi)^2\epsilon}
  \left[\frac43-\alpha_{\rm R}-\frac3{\alpha_{\rm R}}
  \right],
\label{eq:Z_alpha}
\end{equation}
\begin{equation}
Z_g^{(1)}
 =-\frac12Z_a^{(1)}
 =-\frac{11}3\frac{(\mu^{-\epsilon}g_{\rm R})^2}{(4\pi)^2\epsilon},
\label{eq:Z_g}
\end{equation}
where $\epsilon$ is defined as $\epsilon:=(4-d)/2$.

In this paper, we determine the remaining renormalization factors,
$Z_c$, $Z_{\bar c}$ and $Z_C$,
by making use of the dimensional regularization.
In order to determine these three factors,
we must calculate three independent coefficients
$\delta$s in Eqs.~(\ref{eq:deltas}).
For instance, $Z_C$ is obtained by calculating $\delta_a$
in Eq.~(\ref{eq:deltas}).
By calculating $\delta_b$, we obtain a relation of $Z_c$ and $Z_{\bar c}$.
One more relation is obtained by calculating $\delta_e$.
In the actual calculations,
it is useful to remember the fact that the diagonal ghost
does not appear in the internal line.

First, we consider the quantum correction to the diagonal ghost propagator.
There is no divergent graph for the diagonal ghost propagator
in the dimensional regularization,
so that we immediately obtain a relation
between $Z_c^{(1)}$ and $Z_{\bar c}^{(1)}$:
\begin{equation}
\delta_b
 =\frac12Z_c^{(1)}+\frac12Z_{\bar c}^{(1)}
 =0.
\label{eq:relation1}
\end{equation}
Here, the identity $\delta_b=0$ holds to all order of perturbative
calculations, since there is no interaction term including the diagonal
ghost in our action (\ref{eq:S_kappa3}).
In fact, we cannot write the diagram
corresponding to the process which causes radiative corrections to the
diagonal ghost propagator.
In other words, the contribution to the diagonal ghost propagator comes
from a tree level graph alone.

\begin{figure}[tbp]
\begin{center}
\begin{picture}(5400,800)
\put(0,650){\mbox{(a1)}}%
\put(0,0){%
   \put(0,0){\epsfxsize=30mm \epsfbox{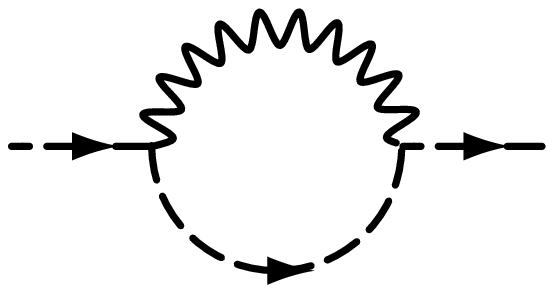}}%
   }%
\put(1400,650){\mbox{(a2)}}%
\put(1400,0){%
   \put(0,0){\epsfxsize=30mm \epsfbox{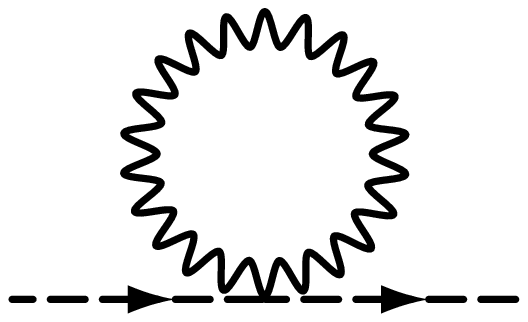}}%
   \put(550,470){\mbox{3}}%
   }%
\put(2800,650){\mbox{(a3)}}%
\put(2800,0){%
   \put(0,0){\epsfxsize=30mm \epsfbox{cc2.eps}}%
   }%
\put(4200,650){\mbox{(a4)}}%
\put(4200,0){%
   \put(0,0){\epsfxsize=30mm \epsfbox{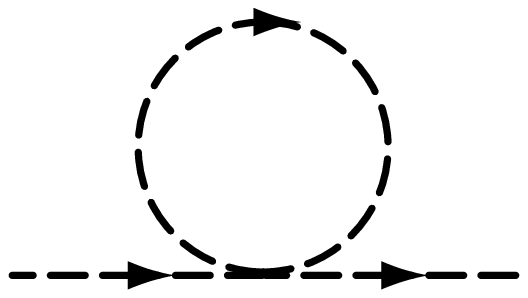}}%
   }%
\end{picture}
\caption[]{%
   The graphs corresponding to one-loop radiative corrections for the propagator of the off-diagonal ghost.
   The wavy line labeled by 3 represents the diagonal gluon
   while the wavy line without any label represents the off-diagonal gluon.
   Similarly the broken line with no label represents the
   off-diagonal ghost or antighost.
}
\label{fig:off-diagonal ghost propagators}
\end{center}
\end{figure}

Next, we consider the quantum correction to 
the off-diagonal ghost propagators.
The divergent graphs for the off-diagonal ghost propagators
are enumerated in Fig.~{\ref{fig:off-diagonal ghost propagators}}.
The graph~(a1) includes both the quadratic and
logarithmic divergences.
On the other hand, each graph~(a2), (a3) and (a4) includes only
quadratic divergence in the dimensional regularization.
However, the quadratic divergence from four graphs are canceled
so that non-trivial (logarithmic) contribution comes
from only one graph~(a1).
Thus, by making use of the dimensional regularization,
$\delta_a$ or $Z_C^{(1)}$ is obtained as
\begin{equation}
\delta_a=Z_C^{(1)}
 =\frac{(g_{\rm R}\mu^{-\epsilon})^2}{(4\pi)^2\epsilon}
  (3-\beta_{\rm R}).
\label{eq:Z_C}
\end{equation}

\begin{figure}[tbp]
\begin{center}
\begin{picture}(2500,1000)
\put(0,850){\mbox{(b1)}}%
\put(0,0){%
   \put(0,0){\epsfxsize=30mm \epsfbox{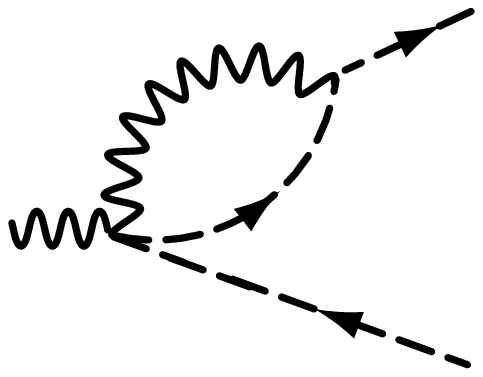}}%
   \put(970,870){\mbox{3}}%
   }%
\put(1400,850){\mbox{(b2)}}%
\put(1400,0){%
   \put(0,0){\epsfysize=1in \epsfbox{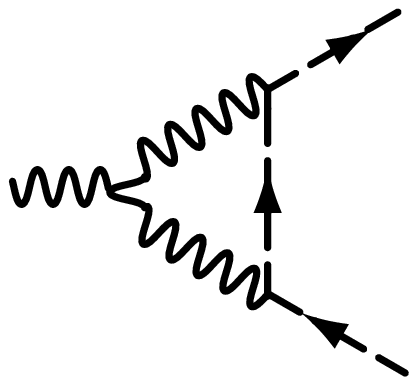}}%
   \put(970,770){\mbox{3}}%
   \put(370,170){\mbox{3}}%
   }%
\end{picture}
\caption[]{%
   The  graphs corresponding to one-loop radiative corrections for the three point vertex of
   one diagonal antighost, one off-diagonal ghost
   and one off-diagonal gluon.
   The wavy line and the broken line labeled by 3
   represent the diagonal gluon and diagonal ghost respectively,
   while the line without any label represents the off-diagonal gluon
   or off-diagonal ghost.
}
\label{fig:3 point vertex}
\end{center}
\end{figure}

In order to determine $\delta_e$ we calculate the quantum correction
to the three point vertex of one diagonal antighost,
one off-diagonal ghost and one off-diagonal gluon.
The divergent graphs for this vertex are collected
in Fig.~{\ref{fig:3 point vertex}}.
Then we obtain
\begin{eqnarray}
\delta_e
 &=&
    \frac12Z_{\bar c}^{(1)}
    +Z_g^{(1)}
    +\frac12Z_A^{(1)}
    +\frac12Z_C^{(1)}
    \nonumber\\
 &=&-\frac{(g_{\rm R}\mu^{-\epsilon})^2}{(4\pi)^2\epsilon}
     \left[\beta_{\rm R}+\frac94+\frac34\alpha_{\rm R}\right],
\end{eqnarray}
or, by solving with respect to $Z_c$ and $Z_{\bar c}$, we also obtain
\begin{equation}
Z_c^{(1)}
 =-Z_{\bar c}^{(1)}
 =(3+\alpha_{\rm R})
   \frac{(g_{\rm R}\mu^{-\epsilon})^2}{(4\pi)^2\epsilon},
\label{eq:Z_kappa}
\end{equation}
where we have made use of Eqs.~(\ref{eq:relation1}), (\ref{eq:Z_A}),
(\ref{eq:Z_g}) and (\ref{eq:Z_C}).

Substituting one loop renormalization factors $Z_X^{(1)}$
and $Z_Y^{(1)}$ into the definitions of
the anomalous dimension~(\ref{eq:def. of anomalous dimension 1})
or RG functions~(\ref{eq:def. of anomalous dimension 2}),
we obtain the following anomalous dimensions and RG functions:
\begin{equation}
\gamma_a(g_{\rm R})
 =-\frac{22}3\frac{g_{\rm R}^2}{(4\pi)^2},
\end{equation}
\begin{equation}
\gamma_A(g_{\rm R})
 =-\frac{g_{\rm R}^2}{(4\pi)^2}
   \left[\frac{17}6-\frac{\alpha_{\rm R}}2-\beta_{\rm R}
   \right],
\end{equation}
\begin{equation}
\gamma_{\beta}(g_{\rm R})
 =\frac{44}3\beta_{\rm R}\frac{g_{\rm R}^2}{(4\pi)^2},
\end{equation}
\begin{equation}
\gamma_\alpha(g_{\rm R})
 =\frac{g_{\rm R}^2}{(4\pi)^2}
  \left[\frac83\alpha_{\rm R}-2\alpha_{\rm R}^2-6
  \right],
\end{equation}
\begin{equation}
\beta(g_{\rm R})
 =\gamma_g(g_{\rm R})
  =-\frac{22}3\frac{g_{\rm R}^3}{(4\pi)^2},
\end{equation}
\begin{equation}
\gamma_c(g_{\rm R})
 =-(3+\alpha_{\rm R})
   \frac{g_{\rm R}^2}{(4\pi)^2}
 =-\gamma_{\bar c}(g_{\rm R}),
\end{equation}
\begin{equation}
\gamma_C(g_{\rm R})
 =-\frac{g_{\rm R}^2}{(4\pi)^2}
  (3-\beta_{\rm R}).
\end{equation}
Thus we have obtained anomalous dimensions of all the fields and
RG functions of all parameters
at the one-loop level of perturbative expansion based on the dimensional
regularization.

\subsection{Scheme $\rm I\! I$}
\label{sec:kappa}
In the previous scheme,
we absorbed $\kappa$ by a rescaling~(\ref{eq:absorption of kappa}).
However, we can also leave this parameter explicitly.
Especially, in the case of $\kappa=0$, we cannot perform
such a rescaling~(\ref{eq:absorption of kappa}).
Here, some of Feynman rules enumerated in the previous subsection
must be modified as follows.
\begin{enumerate}
\item[(a)${}^\prime$] Diagonal gluon propagator:
\begin{equation}
iD_{\mu\nu}
  =-\frac i{p^2}
    \left[g_{\mu\nu}
          -\left(1-\frac\beta{\kappa^2}\right)
           \frac{p_\mu p_\nu}{p^2}\right].
\end{equation}

\item[(c)${}^\prime$] Diagonal ghost propagator:
\begin{equation}
i\Delta
  =-\frac1{\kappa p^2}.
\end{equation}

\item[(g)${}^\prime$]
One off-diagonal gluon,
one off-diagonal ghost and one diagonal antighost:
\begin{equation}
i\left<\bar C^3(p)C^b(q)A_\mu^c\right>_{\rm bare}
  =-i\kappa g\epsilon^{cb}p_\mu.
\end{equation}
\end{enumerate}
Due to the existence of $\kappa$ and
the renormalization of $\kappa$, we can equate
the renormalization factor of $C^3$ and $\bar C^3$,
i.e., $Z_c=Z_{\bar c}$.
Therefore we can redefine the renormalized diagonal antighost field
and a parameter $\kappa$:
\begin{equation}
\bar C^3=Z_c^{1/2}\bar C^3_{\rm R},
\quad
\kappa=Z_\kappa\kappa_{\rm R}.
\end{equation}
By straightforward calculations, we obtain the RG functions of $\kappa$
and $\hat\beta:=\beta/\kappa^2$ as
\begin{equation}
\gamma_\kappa(g_{\rm R})
 =-2(3+\alpha_{\rm R})\kappa_{\rm R}
   \frac{g_{\rm R}^2}{(4\pi)^2},
\label{eq:gamma_kappa}
\end{equation}
\begin{equation}
\gamma_{\hat\beta}(g_{\rm R})
 =\frac{44}3\hat\beta_{\rm R}\frac{g_{\rm R}^2}{(4\pi)^2}.
\end{equation}
The beta-function, the other anomalous dimensions and RG functions are
identical to those calculated in the previous subsection.
From (\ref{eq:gamma_kappa}), we notice that a fixed point of $\kappa$
exists at $\kappa=0$.
After setting $\kappa=0$, we can obtain a simpler action:
\begin{eqnarray}
S_{\rm GF}
 &=&\int d^4x
    \biggl\{
    -\frac1{2\alpha}(D^\mu A_\mu^a)^2
    -\frac1{2\hat\beta}(\partial^\mu a_\mu)^2
    \nonumber\\
 & &+i\bar C^aD^2C^b
    -ig^2\epsilon^{ad}\epsilon^{cb}\bar C^aC^bA^{\mu c}A_\mu^d
    +\frac\alpha4g^2\epsilon^{ab}\epsilon^{cd}\bar C^a\bar C^bC^cC^d
    \biggr\}.
\label{eq:S_mMA+Abel}
\end{eqnarray}
The off-diagonal part is equal to the modified MA gauge term
and the diagonal part is nothing but the Lorentz gauge for Abelian theory.
It is remarkable that this gauge fixing term has FP conjugation invariance
and $SL(2,R)$ symmetry
for the multiplet of ghost and antighost fields.

\section{Conclusion and discussion} %
In this paper, we have investigated how to construct the most general and 
renormalizable MA gauge for $SU(2)$ Yang-Mills theory and performed 
perturbative calculations for a simplest version of
the $SU(2)$ Yang-Mills theory in the MA gauge.

First, we have constructed the most general gauge fixing term
with the BRST symmetry and the global $U(1)$ symmetry,
but without the global $SU(2)$ symmetry.
By definition, the modified MA gauge partially fixes the gauge symmetry
so that the residual $U(1)$ gauge symmetry leaves intact.
In order to make the gauge fixing term renormalizable in the exact sense, 
furthermore,
we must fix the residual $U(1)$ gauge symmetry and hence
  we need the gauge fixing term for fixing the residual $U(1)$ symmetry. In 
the MA gauge, only the global $U(1)$ symmetry remains unbroken.

Second, we have required several symmetries
in order to restrict the parameter space.
We expect that the renormalizability is not spoiled
by restricting the parameter space to a subspace protected by the imposed 
symmetries.
We have found
that at least three independent parameters $\alpha$, $\beta$
and $\kappa$ are necessary and sufficient
to maintain the renormalizability.
The minimal choice coincides with the modified MA gauge proposed in the 
previous papers \cite{KondoII,KS00a} from the viewpoint of renormalizability.

Third, we have calculated the beta function,
anomalous dimensions of all fields
and RG functions of all gauge parameters in the minimal action
at one loop order.

In the construction of action,  we should not impose so many restrictions.
For instance, the ordinary Faddeev-Popov term
   $i\partial^\mu\bar C^A{\cal D}_\mu C^A$ is not invariant under
   the translation of the diagonal ghost,
   while it is invariant under
   the translation of the diagonal antighost.
Thus, if we would like to compare the MA gauge
   with the ordinary Lorentz gauge,
   we should not require the translational symmetry for the diagonal ghost.

Similarly,  in the  modified MA gauge,
   we expect that the ghost--antighost composite operators
   $\epsilon^{ab}C^a\bar C^b$ and $C^a\bar C^a$ have non-trivial
   expectation values due to the condensation and hence
   the charge conjugation symmetry breaks down.\cite{KS00a,Schaden}
Therefore we must not require the charge conjugation symmetry from this 
viewpoint.
The results of investigations to remedy these shortcomings
will be reported elsewhere.

In this paper, we have used the full BRST transformation with nilpotency
for constructing the MA gauge fixing.  The GF+FP term obtained in this way
contains the diagonal ghost and antighost for $\kappa\not=0$.  In a special
case $\kappa=0$, the diagonal ghost and antighost decouple from the
Lagrangian, leaving only the simple conventional gauge fixing term for the
diagonal gluon, see eq.~(\ref{eq:S_mMA+Abel}).
At $\kappa=0$, the GF+FP term does not
contain the diagonal ghost and antighost and reproduces the previous
version of MA gauge \cite{KS00a}.
If the diagonal ghost is not contained in the MAG Lagrangian from the
beginning, the partition function vanishes, since the measure contains the
diagonal ghost, whereas the functional does not contain it.
In this case,  the correlation function must be obtained by differentiating
$\log Z[J]/Z[0]$ with respect to $J$.  Even if $Z[J]$ contains such a zero,
the normalization $Z[0]$ carries the same zero.  Therefore, the ratio is
still well-defined and hence one can forget about zero.  This is equivalent
to removing the diagonal ghost from the measure.
In fact, Schaden \cite{Schaden} has used a new equivariant BRST which
contains no diagonal ghost from the beginning.  In other words, he
considered the BRST transformation on the functional subspace invariant
under the residual gauge transformation.  However, the equivariant BRST is
not nilpotent.  This may cause the difficulty to the renormalizability
argument.  Nevertheless, the final form of the Lagrangian he obtained is
the same as ours set at $\kappa=0$.
Schaden's consideration will be valid, as far as one does not perform the
gauge fixing for the residual Abelian gauge.
In this paper, the MA gauge without diagonal ghost and antighost used so
far is obtained as a special case $\kappa=0$ of our formulation which
includes also the gauge fixing for the residual Abelian gauge without
losing the nilpotency of the BRST transformation.
This is an advantage of our formulation given in this paper.

The results obtained in this paper are the first step
toward the non-perturbative study of  the low  energy physics,
despite that they were calculated by means of perturbative way,
since the high and low energy region of QCD are closely related to
each other according to renormalization group equation
and analyticity.\cite{OZ80}

\section*{Acknowledgements} %
The authors would like to thank Takeharu Murakami for helpful discussion.

\appendix
\section{Rescaling of the fields preserving BRST transformation
         and its connection to the renormalization}%
\label{sec:rescale}
First, we consider the gauge fixing term
with global $SU(2)$ gauge symmetry given by
\begin{equation}
{\cal L}_{\rm GF}
 =-i\gamma
   \mbox{\boldmath$\delta$}_{\rm B}
     \left[\bar{\cal C}^A
           \left(\partial^\mu{\cal A}_\mu^A
                 +\frac\alpha2{\cal B}^A
                 -\zeta gf^{ABC}\bar{\cal C}^B{\cal C}^C
                 \right)\right],
\label{eq:GF with SU2}
\end{equation}
where $\gamma$, $\alpha$ and $\zeta$ are arbitrary parameters and
$\mbox{\boldmath$\delta$}_{\rm B}$ is the BRST transformation:
\begin{equation}
\begin{array}{l}
\mbox{\boldmath$\delta$}_{\rm B}{\cal A}_\mu^A
 ={\cal D}_\mu{\cal C}^A
 =\partial_\mu{\cal C}^A
  +gf^{ABC}{\cal A}_\mu^B{\cal C}^C,\\[1mm]
\displaystyle
\mbox{\boldmath$\delta$}_{\rm B}{\cal C}^A
 =-\frac g2f^{ABC}{\cal C}^B{\cal C}^C,\\[3mm]
\mbox{\boldmath$\delta$}_{\rm B}\bar{\cal C}^A
 =i{\cal B}^A,\\[1mm]
\mbox{\boldmath$\delta$}_{\rm B}{\cal B}^A
 =0.
\end{array}
\label{eq:BRST}
\end{equation}
Because of the nilpotency of the BRST transformation,
it is trivial that (\ref{eq:GF with SU2}) is invariant
under the transformation:
\begin{equation}
\delta{\cal A}_\mu^A
 =\epsilon\mbox{\boldmath$\delta$}_{\rm B}
  {\cal A}_\mu^A,
\quad
\delta{\cal C}^A
 =\epsilon\mbox{\boldmath$\delta$}_{\rm B}
  {\cal C}^A,
\quad
\delta\bar{\cal C}^A
 =\epsilon\mbox{\boldmath$\delta$}_{\rm B}
  \bar{\cal C}^A,
\quad
\delta{\cal B}^A
 =\epsilon\mbox{\boldmath$\delta$}_{\rm B}
  {\cal B}^A,
\end{equation}
for arbitrary Grassmann parameter $\epsilon$.

Now we show that the parameter $\gamma$ can be set equal to 1
without losing generality.
It turns out that such a parameter can be absorbed by rescaling the fields
${\cal C}^A$ and $\bar{\cal C}^A$.
Indeed, by rescaling the fields as
\begin{equation}
{\cal A}_\mu^\prime
  =x{\cal A}_\mu,
\quad
{\cal B}_\mu^\prime
  =y{\cal B}_\mu,
\quad
{\cal C}^\prime
  =u{\cal C},
\quad
\bar{\cal C}^\prime
  =v\bar{\cal C},
\label{eq:rescaled fields}
\end{equation}
the BRST transformation (\ref{eq:BRST}) is rewritten as
\begin{equation}
\begin{array}{l}
\mbox{\boldmath$\delta$}_{\rm B}{\cal A}_\mu^{\prime A}
 =\left(\frac xu\right)\partial_\mu{\cal C}^{\prime A}
  +\left(\frac1u\right)
   gf^{ABC}{\cal A}_\mu^{\prime B}{\cal C}^{\prime C},\\[2mm]
\mbox{\boldmath$\delta$}_{\rm B}{\cal C}^{\prime A}
 =-\left(\frac1u\right)
   \frac12gf^{ABC}{\cal C}^{\prime B}{\cal C}^{\prime C},\\[2mm]
\mbox{\boldmath$\delta$}_{\rm B}\bar{\cal C}^{\prime A}
 =i\left(\frac vy\right){\cal B}^{\prime A},\\[2mm]
\mbox{\boldmath$\delta$}_{\rm B}{\cal B}^{\prime A}
 =0.
\end{array}
\end{equation}
If two conditions $x=1$ and $y=uv$ are satisfied,
the same form of the BRST transformation
as the original BRST transformation~(\ref{eq:BRST}) is obtained
for the rescaled field~(\ref{eq:rescaled fields})
by defining a new BRST transformation
$\mbox{\boldmath$\delta$}_{\rm B}^\prime
  :=u\mbox{\boldmath$\delta$}_{\rm B}$
as
\begin{equation}
\begin{array}{l}
\mbox{\boldmath$\delta$}_{\rm B}^\prime{\cal A}_\mu^{\prime A}
 =\partial_\mu{\cal C}^{\prime A}
  +gf^{ABC}{\cal A}_\mu^{\prime B}{\cal C}^{\prime C},\\[2mm]
\mbox{\boldmath$\delta$}_{\rm B}^\prime{\cal C}^{\prime A}
 =-\frac g2f^{ABC}{\cal C}^{\prime B}{\cal C}^{\prime C},\\[2mm]
\mbox{\boldmath$\delta$}_{\rm B}^\prime\bar{\cal C}^{\prime A}
 =i{\cal B}^{\prime A},\\[2mm]
\mbox{\boldmath$\delta$}_{\rm B}^\prime{\cal B}^{\prime A}
 =0.
\end{array}
\end{equation}
Then the Lagrangian~(\ref{eq:GF with SU2}) is rewritten as
\begin{equation}
{\cal L}_{\rm GF}
 =-i\gamma^\prime
   \mbox{\boldmath$\delta$}_{\rm B}^\prime
     \left[\bar{\cal C}^{\prime A}
           \left(\partial^\mu{\cal A}_\mu^{\prime A}
                 +\frac{\alpha^\prime}2{\cal B}^{\prime A}
                 -\zeta^\prime
                  gf^{ABC}\bar{\cal C}^{\prime B}{\cal C}^{\prime C}
                 \right)\right],
\end{equation}
where $\gamma^\prime:=\gamma/y$, $\alpha^\prime:=\alpha/y$ and
$\zeta^\prime:=\zeta/y$.
Therefore we can set $\gamma^\prime$ to 1 by requiring a condition
$y=\gamma$.

Here, we have introduced four rescaling parameters ($x$, $y$, $u$ and $v$)
and imposed three conditions ($x=1$, $y=uv$ and $\gamma=y$).
Then, we can require one more condition.
We consider the renormalization of fields and parameter as
\begin{equation}
{\cal C}=Z_C^{1/2}{\cal C}_{\rm R},\quad
\bar{\cal C}=Z_{\bar C}^{1/2}\bar{\cal C}_{\rm R},\quad
u=Z_uu_{\rm R},\quad
v=Z_vv_{\rm R},
\label{eq:renormalization1}
\end{equation}
while
\begin{equation}
{\cal C}^\prime=Z_{C^\prime}^{1/2}{\cal C}_{\rm R}^\prime,\quad
\bar{\cal C}^\prime=Z_{\bar C^\prime}^{1/2}\bar{\cal C}_{\rm R}^\prime.
\label{eq:renormalization2}
\end{equation}
In general, it is not necessary that the renormalization factors
$Z_C$ and $Z_{\bar C}$
(similarly $Z_{C^\prime}$ and $Z_{\bar C^\prime}$) are equivalent.
However, substituting~(\ref{eq:renormalization1})
and  (\ref{eq:renormalization2}) into (\ref{eq:rescaled fields}),
we have the relations:
\begin{equation}
Z_{C}^{1/2}Z_u{\cal C}_{\rm R}u_{\rm R}
 =Z_{C^\prime}^{1/2}{\cal C}_{\rm R}^\prime
\quad\mbox{and}\quad
Z_{\bar C}^{1/2}Z_v\bar{\cal C}_{\rm R}v_{\rm R}
 =Z_{\bar C^\prime}^{1/2}\bar{\cal C}_{\rm R}^\prime,
\end{equation}
and we can require the relation $Z_{C^\prime}=Z_{\bar C^\prime}$
by taking $u$ and $v$ appropriately.
Therefore we adopt it as the last condition.

Next, we consider the case of the gauge fixing term with global U(1)
gauge symmetry alone in $SU(2)$ Yang-Mills theory.
Such a term is given by
\begin{equation}
{\cal L}_{\rm GF}
 :=i\gamma\partial^\mu\bar C^a\partial_\mu C^a
   +i\kappa\partial^\mu\bar C^3\partial_\mu C^3
   +\cdots,
\end{equation}
where ``$\cdots$'' denotes the interaction terms given
in section~\ref{sec:The most general gauge fixing terms}.
Decomposing the BRST transformation~(\ref{eq:BRST}) into
diagonal and off-diagonal components explicitly, we obtain
\begin{equation}
\begin{array}{ll}
\mbox{\boldmath$\delta$}_{\rm B}a_\mu
 =\partial_\mu C^3
    +g\epsilon^{ab}A_\mu^aC^b,
&\mbox{\boldmath$\delta$}_{\rm B}A_\mu^a
 =\partial_\mu C^a
    +g\epsilon^{ab}A_\mu^bC^3
    -g\epsilon^{ab}a_\mu C^b,\\[2mm]
\mbox{\boldmath$\delta$}_{\rm B}C^3
 =-\frac12g\epsilon^{ab}C^aC^b,
&\mbox{\boldmath$\delta$}_{\rm B}C^a
 =-g\epsilon^{ab}C^bC^3,\\[2mm]
\mbox{\boldmath$\delta$}_{\rm B}\bar C^3
 =iB^3,
&\mbox{\boldmath$\delta$}_{\rm B}\bar C^a
 =iB^a,\\[1mm]
\mbox{\boldmath$\delta$}_{\rm B}B^3
 =0,
&\mbox{\boldmath$\delta$}_{\rm B}B^a
 =0.
\end{array}
\label{eq:BRST U1}
\end{equation}
After rescaling the fields as
\begin{equation}
\begin{array}{llll}
a_\mu^\prime=ka_\mu,
&B^{\prime3}=lB^3,
&C^{\prime3}=mC^3,
&\bar C^{\prime3}=n\bar C^3,\\[2mm]
A_\mu^{\prime a}=xA_\mu^a,
&B^{\prime a}=yB^a,
&C^{\prime a}=uC^a,
&\bar C^{\prime a}=v\bar C^a,
\end{array}
\end{equation}
the BRST transformation is rewritten as
\begin{equation}
\begin{array}{l}
\ \mbox{\boldmath$\delta$}_{\rm B}a_\mu^\prime
 =\left(\frac km\right)
  \partial_\mu C^{\prime3}
  +\left(\frac k{xu}\right)
   g\epsilon^{ab}A_\mu^{\prime a}C^{\prime b},\\[2mm]
\ \mbox{\boldmath$\delta$}_{\rm B}A_\mu^{\prime a}
 =\left(\frac xu\right)
  \partial_\mu C^{\prime a}
  +\left(\frac1m\right)
   g\epsilon^{ab}A_\mu^{\prime b}C^{\prime3}
  -\left(\frac x{ku}\right)
   g\epsilon^{ab}a_\mu^\prime C^{\prime b},\\[2mm]
\begin{array}{ll}
\mbox{\boldmath$\delta$}_{\rm B}C^{\prime3}
 =-\left(\frac m{u^2}\right)
   \frac12g\epsilon^{ab}C^{\prime a}C^{\prime b},
&\mbox{\boldmath$\delta$}_{\rm B}C^{\prime a}
 =-\left(\frac1m\right)
   g\epsilon^{ab}C^{\prime b}C^{\prime3},\\[2mm]
\mbox{\boldmath$\delta$}_{\rm B}\bar C^{\prime3}
 =\left(\frac nl\right)
   iB^{\prime3},
&\mbox{\boldmath$\delta$}_{\rm B}\bar C^{\prime a}
 =\left(\frac vy\right)
   iB^{\prime a},\\[1mm]
\mbox{\boldmath$\delta$}_{\rm B}B^{\prime3}
 =0,
&\mbox{\boldmath$\delta$}_{\rm B}B^{\prime a}
 =0.
\end{array}
\end{array}
\end{equation}
Similarly to the previous case, imposing the conditions:
\begin{equation}
\frac km
 =\frac k{xu}
 =\frac xu
 =\frac1m
 =\frac x{ku}
 =\frac m{u^2}
 =\frac nl
 =\frac vy,
\end{equation}
or
\begin{equation}
k=1,\quad
x^2=1,\quad
m^2=u^2,\quad
l=mn,\quad
y=mv,
\label{eq:5conditions}
\end{equation}
we can obtain the some BRST transformation as (\ref{eq:BRST U1})
for the rescaled fields:
\begin{equation}
\begin{array}{ll}
\mbox{\boldmath$\delta$}_{\rm B}^\prime a_\mu^\prime
 =\partial_\mu C^{\prime3}
    +g\epsilon^{ab}A_\mu^{\prime a}C^{\prime b},
&\mbox{\boldmath$\delta$}_{\rm B}^\prime A_\mu^{\prime a}
 =\partial_\mu C^{\prime a}
    +g\epsilon^{ab}A_\mu^{\prime b}C^{\prime3}
    -g\epsilon^{ab}a_\mu^\prime C^{\prime b},\\[2mm]
\mbox{\boldmath$\delta$}_{\rm B}^\prime C^{\prime3}
 =-\frac12g\epsilon^{ab}C^{\prime a}C^{\prime b},
&\mbox{\boldmath$\delta$}_{\rm B}^\prime C^{\prime a}
 =-g\epsilon^{ab}C^{\prime b}C^{\prime3},\\[2mm]
\mbox{\boldmath$\delta$}_{\rm B}^\prime\bar C^{\prime3}
 =iB^{\prime3},
&\mbox{\boldmath$\delta$}_{\rm B}^\prime\bar C^{\prime a}
 =iB^{\prime a},\\[1mm]
\mbox{\boldmath$\delta$}_{\rm B}^\prime B^{\prime3}
 =0,
&\mbox{\boldmath$\delta$}_{\rm B}^\prime B^{\prime a}
 =0,
\end{array}
\end{equation}
where we have defined the new BRST transformation
$\mbox{\boldmath$\delta$}_{\rm B}^\prime$ as
$\mbox{\boldmath$\delta$}_{\rm B}^\prime
  :=m\mbox{\boldmath$\delta$}_{\rm B}$
by using the rescaling factor $m$ of the diagonal ghost $C^3$.

Here, we have introduced five conditions~(\ref{eq:5conditions})
for eight parameters ($k$, $l$, $m$, $n$, $x$, $y$, $u$, $v$).
Therefore, we can impose three more conditions.
Note that there are four options:
\begin{enumerate}
\item[(i)]
An absorption of a parameter $\gamma$,

\item[(ii)]
An absorption of a parameter $\kappa$,

\item[(iii)]
An equivalence of renormalization factor of $C^a$ and $\bar C^a$,

\item[(iv)]
An equivalence of renormalization factor of $C^3$ and $\bar C^3$.

\end{enumerate}
Since we can impose only three conditions,
one of the four options is never satisfied.
It is possible to discard a condition (iii) or (iv)
as done in Refs.~\cite{MLP85} and \cite{HN93}.
However, in order to deal with the ghost and antighost on equal footing,
for example, FP conjugation or BRST--anti-BRST field formalism,
it is useful to retain the parameter $\gamma$ or $\kappa$
as we have adopted in this paper.

Thus we can restrict the gauge fixing terms with global $U(1)$
without losing generality to
\begin{equation}
{\cal L}_{\rm GF}
 :=i\partial^\mu\bar C^a\partial_\mu C^a
   +i\kappa\partial^\mu\bar C^3\partial_\mu C^3
   +\cdots,
\end{equation}
where the renormalization factors of $\bar C^a$ and $C^a$ are
identical to each other and this is also the case for $\bar C^3$ and $C^3$.
It is remarkable that the parameter $\kappa$ (or $\gamma$)
cannot be absorbed.

\end{document}